\documentstyle[aps,pre,epsfig]{revtex}\twocolumn
\newcommand{\T}{{\cal T}}
\newcommand{\Ord}{{\cal O}}
\newcommand{\xs}{\bar x_s}
\newcommand{\xu}{\bar x_u}
\newcommand{\xsu}{\bar x_{s,u}}
\newcommand{\PA}{P_{\cal A}}
\newcommand{\opt}{{\rm opt}}
\newcommand{\dx}{\Delta x^\ast}
\newcommand{\J}{\Lambda}
\newcommand{\A}{I}
\newcommand{\tc}{t_{1}}

\begin{document}
%\title{Thermally activated escape in a periodically driven potential}
\title{Surmounting Oscillating Barriers: Path-integral approach for Weak Noise}
\author{J\"org Lehmann, Peter Reimann,  and Peter H\"anggi}
\address{Universit\"at Augsburg, 
         Institut f\"ur Physik,
         Universit\"atsstra\ss e~1, 
         D-86135 Augsburg, 
         Germany}
\maketitle
\begin{abstract}
  We consider the thermally activated escape of an overdamped Brownian 
  particle over a potential barrier in the presence of periodic driving. 
  A time-dependent path-integral formalism is developed which allows us to
  derive asymptotically exact weak-noise expressions for both the 
  {\em instantaneous} and the {\em time-averaged} escape rate.  
  Our results comprise a conceptionally new, systematic treatment of the 
  {\em rate prefactor} multiplying the exponentially leading Arrhenius 
  factor. 
  Moreover, an estimate for the deviations at {\em finite} noise-strengths 
  is provided and a supersymmetry-type property of the time averaged
  escape rate is verified.
  For piecewise parabolic potentials, the rate-expression
  can be evaluated in closed analytical form, while
  in more general cases, as exemplified by a cubic potential,
  an action-integral remains to be minimized numerically.
  Our comparison with very accurate numerical results demonstrates
  an excellent agreement with the theoretical predictions
  over a wide range of driving strengths and driving frequencies.
\end{abstract}
\vspace{2mm}

\section{Introduction}
The thermally activated escape over a potential barrier is a recurrent
theme in a large variety of physical, chemical, and biological contexts
\cite{han90,fle93,tal95}.
In the case of foremost practical relevance, the characteristic 
strength of the thermal noise (the thermal energy $k_BT$) is
much smaller than the potential barrier. 
As a consequence, successful barrier crossings constitute rare events
and the escape statistics verifies with very high
accuracy an exponential decay as a function of time.
In other words, a meaningful escape-rate can be defined which 
completely characterizes the decay process.
A seminal contribution to the theory of escape-rates represents
the work by Kramers in 1940 \cite{kra40}, which
subsequently has been refined, modified, and generalized in various
important directions \cite{han90,fle93,tal95}.

A particularly challenging novel direction are systems far 
away from thermal equilibrium, either due to non-thermal 
noise or external deterministic forces \cite{han90}.
In such a case, the relevant probability distribution
strongly deviates from the Boltzmann-form in the entire
state space and its determination becomes a highly 
non-trivial problem.
{\em Mutas mutandis}, 
this very same basic difficulty resurfaces again 
in all known theoretical methods of calculating escape rates
in far from equilibrium systems 
\cite{tal87,dro92,han95,lud75,gra85,wio89,ein95,mai97}.

The subject of our present paper is one of the simplest
non-equilibrium descendants of Kramers' original  
problem: namely the thermally activated escape of an 
overdamped Brownian particle over a potential barrier 
in the presence of a periodic driving (details
are given in Sect.~II).
This is a prototypical setup in the sense that investigating
the behavior of a system under the influence of a periodic 
forcing represents a particularly natural and
straightforward experimental situation.
Examples arise in the context of
laser driven semiconductor heterostructures \cite{kea}, 
stochastic resonance \cite{sr}, 
directed transport in rocked Brownian motors \cite{rat1,rat2,rat3}, 
or periodically driven ``resonant activation'' processes \cite{ra1,ra2}
like AC driven biochemical reactions in 
protein membranes \cite{ser}, to name only a few.

Despite its experimental importance, the theory of
oscillating barrier crossing in the regime of
weak thermal noise is still at its beginning. 
Previous quantitative,
analytical investigations have been restricted 
to weak \cite{gra84,jun93,sme99}, 
slow \cite{jun89,tal99},
or fast \cite{gra84,jun89,rei96a} driving.
In this article we continue our
recent study \cite{leh00}
of the most challenging intermediate regime
of {\it moderately strong} and 
{\em moderately fast} driving
by means of path-integral methods. 
The general framework of this approach is
derived from scratch in Sect.~III, thus
collecting, streamlining, and partially extending
previously known material.
The evaluation of the escape-rate is worked out
in Sect.~IV with the central results (\ref{4.60}) for the
time-averaged and (\ref{4.51}) for the instantaneous escape 
rate. Especially, these results comprise 
a conceptionally new, systematic treatment of the 
{\it rate prefactor} multiplying 
the exponentially leading Arrhenius factor.
They become asymptotically exact 
for any finite amplitude and period of the driving 
as the noise strength tends to zero.
On the other hand, for
any fixed (small) noise strength, 
we have to exclude extremely small driving amplitudes and 
extremely long or short driving
periods since this would lead us
effectively back to an undriven escape problem,
which is not covered by our present approach.
Another situation which is excluded in our theory is
the case of extremely strong driving such that escape
events become possible even in the absence of the 
thermal noise \cite{spa96,sto00}.
Closest in spirit to our methodology is the
recent work  \cite{sme99}, which is restricted, 
however, to the linear response regime (weak driving)
for the exponentially 
leading part (Arrhenius factor) and treats the prefactor by means of 
a matching procedure, involving the barrier region only.
The approximation adopted in that work is complementary 
to ours in that it admits, for a fixed (small) noise 
strength, arbitrarily small driving amplitudes.

In Sect.~V, our analytical predictions are 
verified for the case of sinusoidally rocked 
metastable potentials against very precise numerical results.
A first example consists in a piecewise parabolic potential,
for which our general rate-expressions
can be evaluated in closed analytical form.
In more general cases, exemplified in Sect.~V by a cubic potential,
a few elementary numerically tasks remain before actual numbers can
be obtained from our rate-expressions.
The final conclusions are presented in Sect.~VI.

\section{The escape problem}
\subsection{Model}
We consider the following model for the one-dimensional Brownian motion of 
a particle with coordinate $x(t)$ and mass $m$ under the influence of a 
time-dependent force field $F(x,t)$:
\begin{equation}
m\,\ddot x(t) - F(x(t),t) = -\eta\,\dot x(t) + \sqrt{2\, D}\ \xi (t)\ .
\label{2.1}
\end{equation}
While the left hand side accounts for the dynamics of the isolated particle,
the right hand side models the influence of its thermal environment \cite{han90}
with a viscous friction coefficient $\eta$
and a randomly fluctuating force $\xi (t)$, which is assumed to be unbiased 
Gaussian white noise with correlation
\begin{equation}
\langle\xi(t)\,\xi (t')\rangle = \delta (t-t') \ .
\label{2.2} 
\end{equation}
At thermal equilibrium, the intensity $D$ of the noise is related to the 
temperature $T$ according to the Einstein-relation
$D=\eta\,k_B T$, where $k_B$ is Boltzmann's constant \cite{han90}.
Throughout this paper we will restrict ourselves to the overdamped motion
such that inertia effects $m\,\ddot x(t)$ in (\ref{2.1}) are negligible 
\cite{leh99}.
Choosing the time unit such that $\eta = 1$, the stochastic dynamics takes the
form
\begin{equation}
\dot x(t) = F(x(t),t)+\sqrt{2\, D}\ \xi(t)\ .
\label{2.3}
\end{equation}

The force field $F(x,t)$ in (\ref{2.3}) is assumed to 
derive from a metastable potential 
%$V(x,t)$ as cartooned in fig.1 via
%\begin{equation}
%F(x,t)= - \partial_x V(x,t)
%\label{2.4a}
%\end{equation}
which undergoes an arbitrary periodic modulation in time with period $\T$:
\begin{equation}
F(x,t+\T )=F(x,t)\ .
\label{2.4}
\end{equation}
An example is a metastable static potential $V(x)$ as cartooned in fig.1,
supplemented by an additive sinusoidal driving
\begin{eqnarray}
F(x,t) & = & -V'(x)+A\,\sin(\Omega\, t)\label{2.5}\\
\Omega & = & 2\,\pi/\T\label{2.6}
\end{eqnarray}

Our next assumption is that the deterministic dynamics in (\ref{2.3}) with
$D=0$ exhibits a stable periodic orbit $x_s (t)$ and an unstable periodic
orbit $x_u (t)$ \cite{jun91a}, satisfying
\begin{eqnarray}
\dot x_{s,u}(t) & = & F(x_{s,u}(t),t)\label{2.14a}\\
x_{s,u}(t+\T) & = & x_{s,u}(t )  \ , \label{2.14b}
\end{eqnarray}
where `$s,u$' means that the index may be either `$s$' or `$u$'.
Moreover, every deterministic trajectory is assumed to approach in the
long-time limit either the attractor $x_s(t)$ or to diverge towards $x=\infty$,
except if it starts exactly at the separatrix $x_u(t)$ between those
two basins of attraction. In other words, the metastable potential is required 
not to be rocked too violently such that particles cannot escape deterministically,
i.e. without the
assistance of the random fluctuations in (\ref{2.3}).
It is clear that $x_s(t)$ and $x_u(t)$ must be disjoint and by assuming a second
``attractor'' at $x=\infty$ we have, without loss of generality, implicitly
restricted ourselves to case that
\begin{equation}
x_u(t)>x_s(t)
\label{2.14c}
\end{equation}
for all $t$.
Note that the above requirements do not necessarily exclude the possibility that for
certain times
$t$ the ``instantaneous potential'', from which the force field
$F(x,t)$ derives, does no longer exhibit a potential barrier.

\subsection{Escape rates}
Next, we return to the stochastic dynamics (\ref{2.3}) with a finite but
very small noise-strength $D$ such that a particle $x(t)$ is able to leave
the domain of attraction of the stable periodic orbit $x_s(t)$ and
subsequently disappear towards $x=\infty$ but the typical waiting-time
before such an event occurs is much longer than all characteristic time 
scales of the deterministic dynamics (separation of time scales \cite{han90,jun93,rei98}).
For a quantitative characterization of such escape events, our starting
point is the 
probability distribution $p(x,t)$ of particles which is governed
by the Fokker-Planck equation \cite{ris84}
\begin{equation}
\frac{\partial}{\partial t} \, p(x,t) = 
\frac{\partial}{\partial x}\{ - F(x,t)+D\frac{\partial}{\partial x} \}\, p(x,t)\ .
\label{2.17}
\end{equation}
%where $\partial_t$ and $\partial_x$ denote partial derivatives with respect to
%time and space, respectively.
Once $p(x,t)$ is known,
the population $\PA (t)$ of the time-dependent basin of attraction 
${\cal A}(t):=(-\infty, x_u(t)]$
of the stable periodic orbit $x_s(t)$ follows as
\begin{equation}
\PA (t) = \int_{-\infty}^{x_u(t)} p(x,t)\, dx \ . 
\label{2.18} 
\end{equation}
A suggestive definition of the ``instantaneous rate'' $\Gamma (t)$
is then provided by the relative decrease of this population
per time unit
\begin{equation}
\Gamma (t) := -\dot \PA (t)/\PA (t) \ .
\label{2.19}
\end{equation}
We note that particles which leave the domain of attraction give
rise to a positive contribution to $\Gamma (t)$. There is also a certain
probability that particles from outside this domain recross the separatrix $x_u(t)$,
giving rise to a negative contribution to $\Gamma (t)$. In other words,
eq.(\ref{2.19}) is the net flux of particles (outgoing flux minus back-flux)
across the separatrix $x_u(t)$ in units of the remaining population
$\PA (t)$. By exploiting the deterministic dynamics (\ref{2.14a}) for
$x_u(t)$, the Fokker-Planck equation (\ref{2.17}) for $p(x,t)$, and the
definition (\ref{2.18}) of $\PA (t)$ we can rewrite the instantaneous rate 
(\ref{2.19}) as
\begin{equation}
\Gamma (t) = - \frac{D}{\PA (t)}\, \frac{\partial\, p(x_u(t),t)}{\partial x_u(t)}\ .
\label{2.20}
\end{equation}

Inside the metastable state $x<x_u(t)$ the particle distribution is governed by
intrawell relaxation processes.
For small noise-strengths $D$, their characteristic time scales are well separated
from the typical escape time itself \cite{han90,jun93,rei98}. On this time scale of the
intrawell relaxation, transients die out and the distribution $p(x,t)$ approaches
a quasi-periodic dependence on time $t$. 
More precisely, $p(x,t)/\int_{-\infty}^{x_u(t)}
p(x,t)\, dx$ tends, at least for $x\leq x_u(t)$, towards a time-periodic function
as $t$ grows. The same carries over to the escape probability (\ref{2.20}) and 
thus the time-averaged escape rate
\begin{equation}
\bar\Gamma := \frac{1}{\T}\,\int_{t}^{t+\T}\Gamma(t')\, dt'
\label{2.21}
\end{equation}
becomes independent of the time $t$.

Our assumption of weak noise guarantees that the loss of population $\PA (t)$ is 
negligible on the time scale of the intrawell relaxation for any initial
distribution $p(x,t_0)$ that is negligibly small in the vicinity and
beyond the instantaneous separatrix $x_u(t_0)$. The denominator in (\ref{2.20})
can thus be approximated by $1$ for all times $t-t_0$ much smaller than
the characteristic escape time $1/\bar\Gamma$ itself. Further, we can 
restrict ourselves to delta-distributed initial conditions of the form
$p(x,t_0) = \delta(x-x_0)$ with $x_0$ inside the basin of attraction 
${\cal A}(t)$ of $x_s(t)$
such that the overwhelming majority of realizations (\ref{2.3}) will first
relax towards a close neighborhood of the attractor $x_s(t)$ before they
escape. The behavior of more general initial distributions then readily follows
by way of linear superposition. Moreover, one expects
\cite{han90,jun93,rei98} that after transients (intrawell
relaxation processes) have died out, the time dependent escape rate
(\ref{2.20}) will actually become independent of the initial conditions $x_0$ and $t_0$.
Denoting by $p(x,t \, |\, x_0,t_0)$ the conditional probability associated
with an initial delta peak at $x_0$ we thus can rewrite (\ref{2.20}) as
\begin{equation}
\Gamma (t) = -D\, \frac{\partial\, p( x_u(t),t \, |\, x_0,t_0)}{\partial\, x_u(t)}\ .
\label{2.22}
\end{equation}
We recall that this expression is valid even if $t-t_0$ is not large, but 
then $\Gamma (t)$ still depends on $x_0$ and $t_0$. On the other hand, $t-t_0$ has 
been assumed to be much smaller than the typical escape time $1/ \bar\Gamma$.
However, on this time-scale the rate $\Gamma (t)$ has practically converged
to its asymptotically periodic behavior and thus
the extrapolation of $\Gamma (t)$ to
arbitrarily large $t-t_0$ is trivial.

\subsection{Supersymmetry}
Given a time-periodic force field $F(x,t)$, we define its
supersymmetric partner field $\tilde F(x,t)$ \cite{jun93,jun91,rei00} 
according to
\begin{equation}
\tilde F(x,t) := F(-x,-t )\ .
\label{susy1}
\end{equation}
For instance, if the force field $F(x,t)$ derives from a 
periodically rocked potential $V(x)$ like in (\ref{2.5}),
then its supersymmetric partner is obtained by 
turning $V(x)$ upside down, followed by an inversion of the
$x$-axis, i.e. $\tilde V(x)= - V(-x)$, see fig.2,
while the driving $A\,\sin (\Omega t)$ in (\ref{2.5}) remains
invariant (up to an irrelevant phase).
In such a supersymmetric partner field $\tilde F(x,t)$, the
stable and unstable periodic orbits exchange
their roles, thus defining a new escape problem
out of the basin of attraction of the new stable
orbit $\tilde x_s(t)= -x_{u}(-t)$ across the new separatrix 
$\tilde x_u(t) = -x_s(-t)$.

It has been demonstrated in \cite{jun93,jun91} that for force fields 
$F(x,t)$ of the
form (\ref{2.5}), the time-averaged rate (\ref{2.21})
is invariant under the supersymmetry transformation 
(\ref{susy1}). 
The same line of reasoning \cite{jun93,jun91}
can be readily generalized to force fields of the form
$F(x,t)=-V'(x)+y(t)$ with an arbitrary periodic driving $y(t)$.
In our present paper we will show that for asymptotically
weak noise $D$ 
{\em the time averaged
escape rate (\ref{2.21}) is invariant under the general
supersymmetry transformation (\ref{susy1})
without any further restrictions on $F(x,t)$}.
Regarding the notion of supersymmetry and especially
its connection with supersymmetric quantum mechanics, we refer
to \cite{jun93,jun91} and further references therein.
We finally remark, that the standard definition of the supersymmetric
partner force field is $-F(x,-t)$. For our present purposes,
the definition (\ref{susy1}) is equivalent but more convenient.

\section{Path-integrals: general framework}
In this section the general framework of a path-integral 
approach to the stochastic dynamics (\ref{2.3}) is outlined.
Though these concepts are not new 
\cite{lud75,gra85,wio89,ein95,mai97,fre84,wei79,car81,sch81,lan82,han89,han93}, 
we find it worth while
to briefly review them here in order to make our paper 
self-contained.
We also note that most of this section remains valid beyond the particular assumptions
on the force field $F(x,t)$ from Sect.~II.

\subsection{Time-discretized path-integrals}
Much like in quantum mechanics, also in the present context of stochastic processes,
path-integral concepts have a tangible meaning only when considered as the limiting case
of appropriate discrete-time approximations.
Our first step is therefore a discretization in time of
the overdamped stochastic dynamics (\ref{2.3}).
Denoting the initial and final times by $t_0$ and $t_f$, we introduce the definitions
\begin{eqnarray}
\Delta t & := & [t_f - t_0 ]/N\label{3.1a}\\
t_n & := & t_0+n\,\Delta t\label{3.1b}\\
x_n & := &  x(t_n) \ , \label{3.1c}
\end{eqnarray}
where $n=0,1,...,N$. The integer $N$ is considered as large but finite and will ultimately 
be sent to infinity (continuous time limit). The discretized dynamics (\ref{2.3}) then takes the form
\begin{equation}
x_{n+1} - x_n = F(x_n,t_n)\, \Delta t + \sqrt{2\, D\, \Delta t}\ \xi_n \ 
\label{3.2}
\end{equation}
where the $\xi_n$ are independent, identically distributed Gaussian random numbers with
probability distribution
\begin{equation}
P(\xi_n ) = (2\,\pi)^{-1/2}\ \exp\{ -\xi^2_n/2\} \ .
\label{3.3}
\end{equation}
As a side remark we notice that the so-called ``prepoint discretization scheme''
\cite{lan82,han89,han93} (not to be confused with the Ito-scheme in the stochastic
dynamics (\ref{2.3})) has been implicitly adopted in (\ref{3.2}) for the sake of later 
convenience. Other ``discretization schemes'' 
\cite{lan82,han89,han93} would give rise to a somewhat 
modified path-integral formalism 
but would -- of course -- lead to identical results as far as 
the actual stochastic dynamics (\ref{2.3}) is concerned. 
In passing we further
note that our treatment for Eq.~(\ref{2.3}) can be generalized 
to multiplicative noise $g(x)\,\xi(t)$, with $g(x)\not=0$, 
without encountering additional difficulties. 

For the conditional probability $p_N(x_{n+1}, t_{n+1}\, | \, x_n, t_n )$ to reach the point
$x_{n+1}$ at time $t_{n+1}$ when starting out from $x_n$ at the previous time step $t_n$ 
we find from the discretized dynamics (\ref{3.2}) and the noise-distribution (\ref{3.3}) 
that
\begin{eqnarray}
& & p_N(x_{n+1}, t_{n+1}\, | \, x_n , t_n ) = \nonumber\\  
& &\int  \delta (x_{n+1}  -   x_n   
-   F(x_n, t_n) \Delta t   -   \sqrt{2 D \Delta t}\, \xi_n) 
\, P(\xi_n)\, d\xi_n\nonumber\\
& & = \frac{1}{(4\pi D \Delta t)^{\frac{1}{2}}}\ 
\exp\left\{-\frac{[x_{n+1}-x_n-F(x_n,t_n)\,\Delta t]^2}{4D\Delta t}\right\}\ .
\label{3.4}
\end{eqnarray}
Here and in the following, integrals over the entire real axis are written without
the integration limits $\pm\infty$.
Further, the mutual independence of the random numbers $\xi_n$ in (\ref{3.2})
(Markov property) implies for the conditional probability the Chapman-Kolmogorov relation
\begin{eqnarray}
& & p_N(x_{n+2}, t_{n+2}\, | \, x_n, t_n )=\nonumber\\
& & \int p_N(x_{n+2}, t_{n+2}\, | \, x_{n+1}, t_{n+1} )\ 
p_N(x_{n+1}, t_{n+1}\, | \, x_n, t_n )\ d x_{n+1} \ .\nonumber
%\label{3.5}
\end{eqnarray}
Upon iteration of this relation in combination with (\ref{3.4}) one finds for the
conditional probability the time-discretized path-integral representation
\begin{eqnarray}
& & p_N(x_f, t_f\, | \, x_0, t_0 ) = \nonumber \\
& & \int\frac{dx_1\cdots dx_{N-1}}{(4\,\pi\,D\, \Delta t)^{N/2}}\ 
\exp\left\{-\frac{S_N(x_0,...,x_N)}{D}\right\}\ ,
\label{3.6}
\end{eqnarray}
where
\begin{equation}
S_N(x_0,...,x_N) := \sum_{n=0}^{N-1}
\frac{\Delta t}{4}\, \left[\frac{x_{n+1}-x_n}{\Delta t}- F(x_n, t_n)\right]^2
\label{3.7}
\end{equation}
is the discrete-time ``action'' or ``Onsager-Machlup functional''. 
While $x_1,...,x_{N-1}$ are integration variables in (\ref{3.6}), the initial- 
and end-points are fixed by the prescribed $x_0$ and by the additional constraint
$x_N = x_f$, see (\ref{3.1a})-(\ref{3.1c}).

\subsection{Saddle-point approximation}
For small noise-strengths $D$ the path-integral (\ref{3.6}) is 
dominated by the minima 
of the action $S_N(x_0,...,x_N)$. The existence of at least one (global) 
minimum can be readily inferred from the general structure of
the action in (\ref{3.7}). To keep things simple we assume for the moment that
besides this global minimum no additional (local) minima play a role in (\ref{3.6}).
Denoting the global minimum by ${\bf x} ^\ast := (x_0^\ast,...,x_N^\ast )$
it follows that it satisfies the extremality conditions
\begin{equation}
\frac{\partial S_N({\bf x}^\ast)}{\partial x_n^\ast} =0
\label{3.8}
\end{equation}
for $n=1,...,N-1$, supplemented by the boundary conditions for $n=0,N$:
\begin{equation}
x^\ast_0 = x_0\ \ ,\ \ \  x^\ast_N = x_f \ .
\label{3.7a}
\end{equation}
Under the assumption that the noise-strength $D$ is small, the path-integral in (\ref{3.6})
can be evaluated by means of a saddle point approximation about the minimizing path 
${\bf x}^\ast$ with the result
\begin{equation}
p_N(x_f, t_f\, | \, x_0, t_0 )
=Z_N({\bf x}^\ast )\ e^{-S_N({\bf x}^\ast)/D}\ [1+\Ord (D)]\ .
\label{3.9}
\end{equation}
where the prefactor  $Z_N({\bf x}^\ast )$ is given by a Gaussian integral of the form
\begin{eqnarray}
Z_N({\bf x}^\ast )& & := 
\int \frac{dy_1\cdots dy_{N-1}}{(4\,\pi\,D\,\Delta t)^{N/2}}\nonumber \\
& & \times \exp\left\{ -\frac{1}{2D}\sum_{n,m=1}^{N-1} y_n\,
\frac{\partial^2 S({\bf x}^\ast)}{\partial x_n^\ast\partial x_m^\ast}\,
y_m\right\}\ ,
\label{3.10}
\end{eqnarray}
and where in the order of magnitude expression $\Ord (D)$ only the dependence 
on the noise-strength $D$ is being kept. 
A more detailed quantitative estimate of this correction
$\Ord (D)$ is a difficult, and to our knowledge unsolved task.

The Gaussian integral in (\ref{3.10}) is readily evaluated 
to yield
\begin{equation}
Z_N({\bf x}^\ast ) := 
\left[ 4\,\pi\,D\,\Delta t\, \det\!\left( 2\,\Delta t\,
\frac{\partial^2 S({\bf x}^\ast)}{\partial x_n^\ast
\partial x_m^\ast}\right)\right]^{-\frac{1}{2}} \ ,
\label{3.11}
\end{equation}
where $\det(A_{nm})$ indicates the determinant of an $N-1\times N-1$ matrix with
elements $A_{nm}$.
As demonstrated in Appendix A, the determinant appearing in (\ref{3.11}) can be rewritten 
in the form of a two-step (second order) linear recursion
\begin{eqnarray}
& & \frac{Q^\ast_{n+1}\nonumber-2\, Q^\ast_n - Q^\ast_{n-1}}{\Delta t^2} = \nonumber\\
& & 2\frac{Q^\ast_n\,F'(x^\ast_n,t_n) - Q^\ast_{n-1}\, F'(x^\ast_{n-1},t_{n-1})}{\Delta t}\nonumber\\
& & - Q^\ast_n\, \left[\frac{x_{n+1}^\ast-x^\ast_n}{\Delta t}-F(x^\ast_n,t_n)\right]
F''(x^\ast_n,t_n) \nonumber\\
& & + Q_n^\ast\, F'(x^\ast_n,t_n)^2 - Q_{n-1}^\ast\, F'(x^\ast_{n-1},t_{n-1})^2
\label{3.12}
\end{eqnarray}
with initial conditions
\begin{equation}
Q^\ast_1 = \Delta t\ \ ,\ \ \ \frac{Q^\ast_2 - Q^\ast_1}{\Delta t} = 1 + \Ord (\Delta t )
\label{3.13}
\end{equation}
from which the prefactor $Z_N({\bf x}^\ast ) $ in (\ref{3.11}) follows as
\begin{equation}
Z_N({\bf x}^\ast ) = [4\,\pi\, D\, Q^\ast_N ]^{-1/2} \ .
\label{3.14}
\end{equation}
The fact that ${\bf x}^\ast$ is a minimum of the action (\ref{3.7})
guarantees that $Q^\ast_N > 0 $. Here and in the following we use the
abbreviations
\begin{equation}
F'(x,t):=\frac{\partial F(x,t)}{\partial x}\ \ ,\ \ \ 
\dot F(x,t) := \frac{\partial F(x,t)}{\partial t}
\label{3.14a}
\end{equation}
and bracket-saving expressions like $f(x)^2$ are understood as $[f(x)]^2$.

As we shall see later, we have to leave room for the possibility that even
for small noise-strengths $D$ more than one (global or local) minimum 
of the action (\ref{3.7}) notably 
contributes to the path-integral expression (\ref{3.6}). We label those various 
non-negligible minima ${\bf x}_k^\ast$ by the discrete
index $k$ but leave for the moment the precise set of $k$-values unspecified.
Each of the minimizing paths ${\bf x}_k^\ast$ thus satisfies an extremality condition of the
form (\ref{3.8}). Under the assumption that those minima ${\bf x}_k^\ast$ are well
separated in the $N-1$-dimensional space of all paths $(x_0,...,x_N)$ appearing
in (\ref{3.6}), the saddle point approximation (\ref{3.9}) simply acquires an extra sum
over $k$ with a corresponding extra index $k$ in (\ref{3.10}-\ref{3.14}).
Combining (\ref{3.9}) and (\ref{3.14}) we thus arrive at
\begin{equation}
p_N(x_f, t_f\, | \, x_0, t_0 )
=\sum_k \frac{\ e^{-S_N({\bf x}_k^\ast)/D}}{(4\pi D Q^\ast_{N,k})^{\frac{1}{2}}}\,
[1+\Ord (D)]\ .
\label{3.15}
\end{equation}

\subsection{Continuous-time limit}
Next we turn to the continuous-time 
limit $N\to\infty$, $\Delta t\to 0$ in (\ref{3.1a}).
The continuous-time conditional probability $p (x_f, t_f\, | \, x_0, t_0 )$ when $N\to\infty$ 
in (\ref{3.6}) is symbolically indicated by the path-integral 
expression \cite{fre84}
\begin{equation}
p (x_f, t_f\, | \, x_0, t_0 ) = 
\int\limits_{x(t_0)=x_0}^{x(t_f)=x_f} {\cal D}x(t)\ e^{-S[x(t)]/D} \ ,
\label{3.16}
\end{equation}
where
\begin{equation}
S[x(t)] := \int_{t_0}^{t_f} L(x(t),\dot x(t),t)\, dt
\label{3.17}
\end{equation}
is the continuous-time limit of the action (\ref{3.7}) with
\begin{equation}
L(x,\dot x,t):= \frac{1}{4}\, [\dot x - F(x,t)]^2
\label{3.18}
\end{equation}
as Lagrangian. The extremality conditions for the minimizing paths $x_k^\ast(t)$ 
in the continuous-time limit are obtained from (\ref{3.7}) and (\ref{3.8}) by
letting $\Delta t\to 0$ as
\begin{equation}
\ddot x^\ast_k(t) = \dot F(x^\ast_k(t),t) 
+ F(x^\ast_k(t),t) \, F'(x^\ast_k(t),t)
\label{3.19}
\end{equation}
with boundary conditions (cf. (\ref{3.7a}))
\begin{equation}
x_k^\ast(t_0) = x_0 \ \ ,\ \ \ x_k^\ast(t_f) = x_f \ .
\label{3.20}
\end{equation}
The same result (\ref{3.19}) can also be recovered as the Euler-Lagrange
equation corresponding to the Lagrangian (\ref{3.18}). 

Equivalent to
this Lagrangian dynamics is the following Hamiltonian
counterpart:
\begin{eqnarray}
H(x,p,t) & := & p\, \dot x - L = p^2 + p\, F(x,t)\label{3.21}\\
\dot p^\ast_k(t) & = & -p^\ast_k(t)\, F'(x_k^\ast(t),t)\label{3.22}\\
\dot x^\ast_k(t) & = & 2\, p^\ast_k(t) + F(x_k^\ast(t),t)\ .\label{3.23}
\end{eqnarray}
The last equation (\ref{3.23}) may also be considered as the definition of the momentum
$p^\ast_k (t)$ in terms of $x^\ast_k(t)$ and $\dot x^\ast_k(t)$.
With (\ref{3.18}) and (\ref{3.23}) the action (\ref{3.17}) along a minimizing path 
$x^\ast_k(t)$ follows as
\begin{equation}
\phi_k (x_f, t_f) := S[x_k^\ast(t)] 
= \int_{t_0}^{t_f} p^\ast_k(t)^2\, dt\ ,
\label{3.24}
\end{equation}
where the dependence of the action $\phi_k (x_f, t_f)$
on the initial condition $x_0$ at time $t_0$ has been dropped.
%Here and in the sequel, $p^\ast_k(t)^2$ is a shorthand notation for $[p^\ast_k(t)]^2$.
For later use we also recall the well-known result from classical mechanics
that the derivative of the extremizing action with respect to its endpoint
equals the canonical conjugate momentum, {\em i.e.},
\begin{equation}
\frac{\partial \phi_k (x_f, t_f)}{\partial x_f} = p^\ast_k (t_f)\ .
\label{3.25}
\end{equation}

Finally, the continuous-time limit for the conditional probability
(\ref{3.15}) in combination with (\ref{3.24}) takes the form
\begin{equation}
p (x_f, t_f\, | \, x_0, t_0 )
=\sum_k \frac{\ e^{- \phi_k (x_f, t_f)/D}}
{[4\pi D Q^\ast_{k}(t_f)]^{\frac{1}{2}}}\, [1+\Ord (D)]\ ,
\label{3.26}
\end{equation}
where $Q^\ast_k(t)$ is governed by the second order homogeneous
linear differential equation \cite{leh00,sch81,dre78} that follows 
in the limit $\Delta t\to 0$ from (\ref{3.12}) and (\ref{3.23}):
\begin{eqnarray}
\frac{1}{2}\ddot Q^\ast_k(t) & - & 
\frac{d}{dt} [Q^\ast_k(t)\, F'(x^\ast_k(t),t)]\nonumber\\
& + &  Q^\ast_k(t)\, p^\ast_k(t)\, F''(x^\ast_k(t),t) = 0 \ .
\label{3.27}
\end{eqnarray}
Similarly, the initial conditions (\ref{3.13}) go over 
for $\Delta t\to 0$ into
\begin{equation}
Q^\ast_k(t_0) = 0 \ \ ,\ \ \ \dot Q^\ast_k(t_0) = 1 \ .
\label{3.28}
\end{equation}

We remark that according to (\ref{3.19}) and (\ref{3.20}) the minimizing
paths $x^\ast_k (t)$ are independent of the noise-strength $D$.
Consequently, neither $\phi_k (x_f, t_f)$ from 
(\ref{3.19},\ref{3.20},\ref{3.24}) nor $Q^\ast_k (t)$ from 
(\ref{3.27},\ref{3.28}) depend on the noise-strength, {\em i.e.}, no 
implicit additional $D$-dependences are hidden in (\ref{3.26}).
We further note that by means of the substitution
\begin{equation}
g^\ast_k (t) := \frac{\dot  Q^\ast_k (t)}{2\, Q^\ast_k (t)} - 
F'(x^\ast_k(t),t)
\label{3.29}
\end{equation}
the linear homogeneous second order equation (\ref{3.27}) goes over into the non-linear
first order Riccati-equation
\begin{eqnarray}
\dot g^\ast_k(t) & + &  2\, g^\ast_k(t)^2  +  2\, g^\ast_k(t)\, F'(x^\ast_k(t),t) = \nonumber \\
& - &  p^\ast_k(t)\, F''(x^\ast_k(t),t)\ .
\label{3.30}
\end{eqnarray}
Since (\ref{3.28}) does not lead to a meaningful initial condition for
$g^\ast_k (t)$ in (\ref{3.29}), the Riccati-equation (\ref{3.30}) 
can only be used for times $ t > t_0$.
For this reason and also from the viewpoint of calculational
efficiency we found that for practical purposes
the linear second order equation
(\ref{3.27}) is often superior to the 
Riccati-equation (\ref{3.30}).

To establish contact with previously known results we finally 
remark that one can identify
\begin{equation}
\frac{ \partial^2 \phi_k (x_f, t_f)}{\partial x_f^2} = g^\ast_k(t_f)  \ .
\label{3.31}
\end{equation}
This relation (\ref{3.31}) and the associated Riccati-equation (\ref{3.30})
are usually derived by introducing a WKB-type ansatz into the 
Fokker-Planck-equation for the
conditional probability distribution (cf. (\ref{2.17})) 
and then comparing powers of the
noise-strength $D$. Since a direct derivation by means
of path-integral methods is not known to us, we have included such a 
derivation of the relation (\ref{3.31}) in the Appendix B.

\section{Path-integral solution of the escape problem}
By introducing the path-integral expression (\ref{3.26}) for the conditional
probability into the formula (\ref{2.22}) for the instantaneous rate $\Gamma
(t)$ at time $t=t_f$ and taking into account (\ref{3.25}) 
we obtain {\em our first main result} \cite{leh00}, namely
\begin{equation}
\Gamma (t_f) = 
\sum_k \frac{p^\ast_k (t_f)\, e^{- \phi_k (x_u(t_f), t_f)/D}}
{[4\,\pi\, D\, Q^\ast_{k}(t_f)]^{1/2}}\ [1+\Ord (D)]\ .
\label{4.1}
\end{equation}
In view of (\ref{3.26}), the instantaneous rate (\ref{4.1}) has the
suggestive structure of ``probability at the separatrix times velocity''.

As already mentioned in Sect.~II, for sufficiently large times 
$t_f-t_0$ the instantaneous rate 
(\ref{4.1}) is expected to become independent of the initial position $x_0$
as long as $x_0$ is located inside the domain of attraction of the stable
periodic orbit $x_s(t)$.
A more detailed discussion of this point will be given in Subsect. IV.G. To keep things
as simple as possible we focus in the following subsections on the particular case
that $x_0$ is located {\em at} the stable periodic orbit, {\em i.e.},
\begin{equation}
x_0 = x_s(t_0) \ .
\label{4.2}
\end{equation}

\subsection{Minimizing paths}
Our next goal is the characterization of all the minimizing paths $x^\ast _k(t)$ which
significantly contribute to the sum in (\ref{4.1}). Our first observation is that for 
any finite $t_0$ and $t_f$ the action (\ref{3.17}) exhibits in the generic case a unique 
global minimum respecting the boundary conditions
\begin{equation}
x_k^\ast(t_0) = x_s(t_0)\ \ ,\ \ \ x_k^\ast(t_f) = x_u(t_f)
\label{4.3}
\end{equation}
according to (\ref{3.20}) and (\ref{4.1},\ref{4.2}). To be specific, we denote
this globally minimizing path as $x^\ast_{k_0}(t)$. From the explicit form of
the Lagrangian (\ref{3.18}) we can infer that for large values of $t_f - t_0$
the minimal path $x^\ast_{k_0}(t)$ follows most of the time rather closely a
deterministic trajectory, {\em i.e.}, $\dot x^\ast_{k_0}(t) \simeq
F(x^\ast_{k_0}(t),t)$, in order not to accumulate a too large amount of action
(\ref{3.17}). In view of (\ref{2.14a}) and (\ref{4.3}) it is thus suggestive
that $x^\ast_{k_0}(t)$ starts at $x^\ast_{k_0}(t_0) = x_s(t_0)$ and then
continues to closely follow the stable periodic orbit $x_s(t)$ for quite some
time. At a certain moment, $x^\ast_{k_0}(t)$ leaves this neighborhood and
travels in a comparatively short time into the vicinity of the unstable
periodic orbit $x_u(t)$, where it remains for the rest of its time and ends at
$x^\ast_{k_0}(t_f) = x_u(t_f)$. Only during the crossover from the
neighborhood of $x_s(t)$ into that of $x_u(t)$ does the path $x^\ast_{k_0}(t)$
substantially deviate from a deterministic behavior and so gives rise to the
main contribution to the action (\ref{3.17}).  We desist from a more rigorous
derivation of these basic qualitative features since they are quite similar to
the well-known barrier-crossing problem in a static potential
(time-independent force-field) \cite{wei79,car81,col75}.
Especially, the relatively short ``crossover-segment'' of $x^\ast_{k_0}(t)$
between the long sojourns close to the stable and unstable orbits 
has lead to the name ``instanton'' for such a path.

As we will see in more detail later, a meaningful limit of $x^\ast_{k_0}(t)$
exists for $t_0\to - \infty$ and $t_f\to\infty$ (henceforth abbreviated as
$t_f-t_0\to\infty$) in the sense that $x^\ast_{k_0}(t)$
follows closer and closer the periodic orbits $x_{s,u}(t)$
over longer and longer time intervals,
while the ``crossover-segment'' does practically not change its shape any more.
Also the associated minimal action $S[ x^\ast_{k_0}(t) ]$ from (\ref{3.17}) tends to
a finite limit. In fact, one can readily show that the minimal action cannot increase
upon increasing $t_f$ and/or decreasing $t_0$. Since it is furthermore bounded
from below, the existence of the limit follows for the action as well as for the 
minimizing path itself. More importantly, from the time-periodicity of the force field
(\ref{2.4}) one can infer that in the limit $t_f-t_0\to\infty$
the action $S[ x^\ast_{k_0}(t+n\,\T) ]$ has the same 
value for any integer $n$. In other words, for infinitely large $t_f-t_0$ the action
no longer exhibits a unique absolute minimum, rather each path 
$x^\ast_{k_0}(t+n\,\T )$ globally minimizes the action. However, these
{\em degenerate} absolute minima are still well separated in the space of all paths $x(t)$
appearing in (\ref{3.16}).
This feature is the salient difference between our present 
problem and its time-independent
counterpart \cite{wei79,car81,col75,wei81,wei87}, 
which exhibits a {\em continuous} degeneracy (Goldstone mode)
in the limit $t_f-t_0\to\infty$. Put differently, the time-periodic force field
reduces the continuous time-translation symmetry into a discrete one.
% also on the level of the minimizing paths in the limit $t_f-t_0\to\infty$.
Since the rate-formula (\ref{4.1}) assumes well separated minima 
$x^\ast_k(t)$ of the action,
it is quite clear that the {\em time-independent case must be excluded in the following}.

We emphasize that the minimizing paths $x^\ast_k(t)$ remain 
well separated and thus the rate-formula (\ref{4.1}) 
becomes asymptotically exact for any 
(arbitrary but fixed) finite values of the driving amplitude and period
as the noise strength $D$ tends to zero.
%Formally speaking the $\Ord (D)$ correction 
%of the saddle point approximation (\ref{3.9}) and thus of (\ref{4.1}) 
%remains finite, so 
Apart from this fact that {\em in the limit} $D\to 0$ the $\Ord (D)$ 
correction in the saddle point approximation (\ref{3.9}) and thus in 
(\ref{4.1}) vanishes, a more detailed quantitative statement 
seems difficult.
On the other hand, for
a given (small) noise strength $D$, 
we have to exclude extremely small driving amplitudes and 
extremely long or short driving
periods since this would lead us
effectively back to the static (undriven) escape problem,
which requires a completely different treatment (especially of the 
(quasi-) Goldstone mode \cite{sme99,wei79,car81,col75,wei81,wei87})
than in (\ref{3.9}). 
Put differently, in any of these three asymptotic regimes, 
the error $\Ord (D)$ from (\ref{3.9},\ref{4.1}) 
becomes very large.

For later reference we denote the minimizing path $x^\ast_{k_0}(t)$ when
$t_f - t_0 \to\infty$ by $x^\ast_\opt(t)$, keeping in mind that we are
still free to shift its time argument by an arbitrary multiple of $\T$. The
corresponding action is
\begin{equation}
\phi_\opt := S[x^\ast_\opt(t)] = 
\lim_{
% x(t),\,
         t_{0}\to-\infty \atop t_{f}\to\infty}\!\!\\
\min_{x(t) \atop { x(t_0)=x_s(t_0) \atop  x(t_f)=x_u(t_f)}} \!\! S[x(t)] \ ,
\label{4.4}
\end{equation}
where the second identity may also be considered as an implicit definition of
$x^\ast_\opt(t)$. Similarly, any other quantity associated with $x^\ast_\opt(t)$ will be
marked by an index ``opt'', for instance $p^\ast_\opt(t)$ (see (\ref{3.23})),
$Q^\ast_\opt(t)$ (see (\ref{3.27})), and $g^\ast_\opt(t)$ (see (\ref{3.29})).

In principle, besides the absolute minimum $x^\ast_\opt(t)$ of the action there
may coexist further (absolute or relative) minima which cannot be identified with
each other after a time-shift by an appropriate 
multiple of $\T$. While the coexistence of further
absolute minima is non-generic, coexisting relative minima are irrelevant
for sufficiently small noise-strengths $D$ as far as the sum in (\ref{4.1}) 
is concerned.
Though both cases could be easily taken into account in the following discussion,
we will restrict ourselves to the simplest and most common case that $x^\ast_\opt(t+n\,\T )$ are the
only (relevant)
minima of the action (\ref{3.17}) in the limit $t_f-t_0\to\infty$.

Returning to finite but large values of $t_f-t_0$, we expect -- as a precursor
of the $t_f - t_0 \to\infty$ limit -- that besides the unique absolute minimum
$x^\ast_{k_0}(t)$ there will coexist many additional relative minima
$x^\ast_{k}(t)$ with an only slightly larger action. All those minima $x^\ast_{k}(t)$
possess a limit when $t_f-t_0\to\infty$ in the same sense as for the case $k=k_0$ 
described above (quantitative details will be given later). 
Moreover, when $t_f-t_0\to\infty$ then each $x_k^\ast(t)$ approaches
$x^\ast_\opt(t+n(k)\,\T )$ for a suitable choice of $n(k)$ and without loss
of generality we can assume a (re-)labeling of the $x^\ast_k(t)$ such that $n(k)=k$.
In other words, to each $x^\ast_k(t)$ belongs a 
very similarly looking ``{\em master-path}''
$x^\ast_\opt(t+k\,\T)$, see fig.3.
Since $t_f-t_0$ is finite, there is a finite number (of the order 
$(t_f-t_0)/\T$) of minimizing paths $x^\ast_k(t)$ and without loss of generality we
can assume that the indices in the sum (\ref{4.1}) start at $k=0$ and run until a 
certain maximal value $K(t_f,t_0)$:
\begin{equation}
0\leq k\leq K(t_f,t_0)=\Ord ((t_f-t_0)/\T ) \ .
\label{4.4'}
\end{equation}
Thus $x^\ast_0(t)$ is that minimizing path which closely follows $x_s(t)$ as 
long as possible and
crosses over to the neighborhood of $x_u(t)$ at ``the latest possible moment''
(see fig.3), and similarly for $x^\ast_{K(t_f,t_0)}(t)$.

Note that all the general qualitative features discussed above are nicely
illustrated by the explicit example in Sect.~IV.A.

\subsection{Neighborhood of periodic orbits}
Our final goal is to approximate the action $\phi_k(x_f,t_f)$ and the
prefactor $p^\ast_k(t_f)/[Q^\ast_k(t_f)]^{1/2}$ for all minimizing paths $x^\ast_k(t)$ that play
a non-negligible role in the sum (\ref{4.1},\ref{4.4'}) solely in terms
of the ``master-path'' $x_\opt^\ast(t)$ and its descendents $p^\ast_\opt(t)$,
$Q^\ast_\opt(t)$, etc.
To this end we first address the behavior of a path $x^\ast_k(t)$ within the
neighborhood of either the stable periodic orbit $x_s(t)$ or of the unstable one 
$x_u(t)$. Within these regions,
the time dependent force field can be approximately written as
\begin{equation}
F(x,t)= F(x_{s,u}(t),t) + (x-x_{s,u}(t))\, F'(x_{s,u}(t),t) \ .
\label{4.5}
\end{equation}
Note that these approximations are valid not only if $x-x_{s,u}(t)$ 
is small but also if $F''(y,t)$ is small for all $y$ between $x$ and $x_{s,u}(t)$.
%, {\em e.g.}, for the piecewise parabolic potential (\ref{5.1}) and additive driving (\ref{2.5}).
%The approximation (\ref{4.5}) is furthermore equivalent to its derivatives with respect to $x$:
An immediate consequence of (\ref{4.5}) are the relations
\begin{eqnarray}
F'(x,t)=F'(x_{s,u}(t),t)\ \ ,\ \ \  F''(x,t)=0 \ .
\label{4.6}
\end{eqnarray}
As long as a minimizing path $x^\ast_k(t)$ remains in a region where
these approximations apply, the Hamiltonian equations (\ref{3.22},\ref{3.23}) take the form
\begin{eqnarray}
\dot p^\ast_k(t) & = & - p^\ast_k(t) \, F'(x_{s,u}(t),t)\label{4.7}\\
\Delta\dot x ^\ast_k (t) & = &  2\,p^\ast_k(t) + \dx_k(t)\, F'(x_{s,u}(t),t)\ , 
\label{4.8}
\end{eqnarray}
where we have introduced
\begin{equation}
\dx_k(t) := x^\ast_k(t)-x_{s,u}(t)\ .
\label{4.9}
\end{equation}
Their solutions are:
\begin{eqnarray}
p^\ast_k(t) & = & p^\ast_k(t_1) \, e^{-\J_{s,u}(t,t_1)}\label{4.10}\\
\dx_k(t) & = & \dx_k(t_1)\, 
e^{\J_{s,u}(t,t_1)}+p^\ast_k(t)\, \A_{s,u}(t,t_1)\ ,\label{4.11}
\end{eqnarray}
where $t_1$ is an arbitrary reference time (within our assumption that
(\ref{4.5}) applies for all the considered times $t$) and where
\begin{eqnarray}
\J_{s,u}(t,t_1) & := & \int_{t_1}^t F'(x_{s,u}(t'),t')\, dt'\label{4.12}\\
\A_{s,u}(t,t_1) & := & 2\int_{t_1}^t e^{2\,\J_{s,u}(t,t')}\, dt'\ .\label{4.13}
\end{eqnarray}

Obvious properties of the functions $\J_{s,u}(t,t_1)$ from (\ref{4.12}) are:
\begin{eqnarray}
\J_{s,u}(t,t_2) & = & \J_{s,u}(t,t_1) + \J_{s,u}(t_1,t_2)\label{4.16}\\
\J_{s,u}(t_1,t) & = & - \J_{s,u}(t,t_1)  \label{4.17a}\\
\J_{s,u}(t+\T,t_1+\T) & = & \J_{s,u}(t,t_1) \ . \label{4.18a}
\end{eqnarray}
Further, one readily sees that the quantities
\begin{equation}
\lambda_{s,u} := \J_{s,u}(t+\T,t)/\T
\label{4.17}
\end{equation}
are indeed $t$-independent and characterize the stability/instability
(``Lyapunov exponents'') of the periodic orbits, namely
\begin{equation}
\lambda_s < 0 \ \ , \ \ \ \lambda_u > 0 \ .
\label{4.18}
\end{equation}
One even expects that $\J_s(t,t_1) < 0$ and $\J_u(t,t_1) > 0$  
not only for $t -t_1= \T, 2\T,...$ (cf .(\ref{4.17})) but in fact for all
$t-t_1 > 0$, however exceptions cannot be excluded for not too large $t-t_1$.
 From (\ref{4.12}) and (\ref{4.17}) it follows that $\J_{s,u}(t,t_1)$
can be written as the sum of a linear function $\lambda_{s,u}\cdot (t-t_1)$ 
and a periodic function of $t$. As a consequence, we obtain
\begin{equation}
%\J_s(t,t_1) \stackrel{t\to\infty }{\longrightarrow} - \infty\ \ ,\ \ \ 
%\J_u(t,t_1) \stackrel{t\to\infty }{\longrightarrow} \infty\ .
\J_{s,u}(t,t_1) \sim \lambda_{s,u}\cdot(t-t_1)
\label{4.19}
\end{equation}
for asymptotically large positive and negative times $t-t_1$.

Turning to the discussion of $\A_{s,u}(t,t_1)$ from (\ref{4.13}), 
we first note that
\begin{eqnarray}
\A_s(t,t_1) & = & \A_s(t) - e^{2\,\J_s(t,t_1)}\, \A_s(t_1)\label{4.20}\\
\A_s(t) & := & \lim_{t_1\to - \infty} \A_s(t,t_1) 
= 2 \int\limits_{-\infty}^t e^{2\,\J_s(t,t')}\, dt' \label{4.21}
\end{eqnarray}
and similarly 
\begin{eqnarray}
\A_u(t,t_1) & = & - \A_u(t) + e^{2\,\J_u(t,t_1)}\, \A_u(t_1)\label{4.22}\\
\A_u(t) & := & - \lim_{t_1\to \infty} \A_u(t,t_1) 
= 2 \int\limits_{t}^\infty e^{2\,\J_u(t,t')}\, dt' \ .\label{4.23}
\end{eqnarray}
Thus, $\A_{s,u}(t)$ are positive and finite for all $t$ and obey
\begin{equation}
\A_{s,u}(t+\T)= \A_{s,u}(t )\ .
\label{4.24}
\end{equation}
It follows that $\A_{s}(t,t_1)$ in (\ref{4.20}) is given by a periodic
function of $t$ minus the product of another periodic function times an 
exponentially decreasing factor $\exp\{\lambda_s\cdot (t-t_1)\}$, and
analogously for $\A_u(t,t_1)$ in (\ref{4.22}).

Choosing as reference time
$t_1 =t_0$ in (\ref{4.11}) and taking into account that $\dx_k(t_0)=0$
according to (\ref{4.2},\ref{4.9}), implies that in the neighborhood of 
$x_s(t)$ we have
\begin{equation}
\dx_k(t) = p^\ast_k(t)\, \A_s(t,t_0)\ .
\label{4.25}
\end{equation}
Dividing this result by the same identity evaluated at a different 
reference time $t_s>t_0$ and taking into account (\ref{4.10}) we obtain
\begin{eqnarray}
\dx _k(t) & = & \dx _k(t_s)\, e^{-\J_s(t,t_s)}\,\frac{\A_s(t,t_0)}{\A_s(t_s,t_0)}\label{4.26}\\
p^\ast_k(t) & = & \dx _k(t_s)\, e^{-\J_s(t,t_s)}\, / \, \A_s(t_s,t_0) \ . \label{4.27}
\end{eqnarray}
Both these expressions consist of an exponentially {\em increasing}
factor $\exp\{-\lambda_s\cdot (t-t_s)\}$ times some periodic function of $t$.
In (\ref{4.26}) one has in addition a 
quickly decreasing correction. The corresponding behavior in the neighborhood of $x_u(t)$ 
is given by
\begin{eqnarray}
\dx_k(t) & = & p^\ast_k(t)\, \A_u(t,t_f)\label{4.28}\\
\dx_k(t) & = & \dx_k(t_u)\, e^{-\J_u(t,t_u)}\,\frac{\A_u(t,t_f)}{\A_u(t_u,t_f)}\label{4.29}\\
p^\ast_k(t) & = & \dx_k(t_u)\, e^{-\J_u(t,t_u)}\, / \, \A_u(t_u,t_f) \ ,\label{4.30}
\end{eqnarray}
where $t_u$ is some reference time with $t_u<t_f$. As expected,
(\ref{4.29},\ref{4.30}) are now dominated by an exponentially {\em decreasing}
behavior $\exp\{-\lambda_u\cdot (t-t_u)\}$.
We further remark that for the master-path $x^\ast_\opt(t)$ we have $t_0\to-\infty$ and
$t_f\to\infty$, thus $\A_{s,u}(t,t_{0,f})$ in (\ref{4.25}-\ref{4.30})
go over into $\A_{s,u}(t)$ according to (\ref{4.20},\ref{4.22}) and so 
all four equations
(\ref{4.26},\ref{4.27},\ref{4.29},\ref{4.30}) are {\em exactly} given by 
$\exp\{-\lambda_{s,u}\cdot (t-t_{s,u})\}$ times certain periodic functions of $t$.

Within our above local analysis of the neighborhoods of $x_{s,u}(t)$, 
the reference times $t_{s,u}$ are still arbitrary and the
corresponding parameters
$\dx_k(t_{s,u})$ remain undetermined. They can only be fixed through the global
behavior of $x^\ast_k(t)$. It is instructive to reconsider the same thing from a somewhat different
viewpoint.
 From (\ref{4.20},\ref{4.21},\ref{4.25}) we conclude that
\begin{equation}
\frac{\dx_k(t_s)}{p^\ast_k(t_s)} = \A_s(t_s) - 2\int\limits_{-\infty}^{t_0}e^{2\,\J_s (t_s,t')}\, dt'
\label{4.31}
\end{equation}
and similarly
\begin{equation}
\frac{\dx_k(t_u)}{p^\ast_k(t_u)} = - \A_u(t_u) + 2\int\limits_{t_f}^{\infty}e^{2\,\J_u (t_u,t')}\, 
dt'\ .
\label{4.32}
\end{equation}
Let us consider $t_s>t_0$ as fixed and such that the approximation
(\ref{4.5}) is valid for all $t\in[t_0,t_s]$. Within the same restriction, we now consider the quantity
$\dx_k(t_s)$ as a parameter. For any value of $\dx_k(t_s)$, eq.(\ref{4.31}) thus 
fixes $p^\ast_k(t_s)$. With these initial conditions for $x^\ast_k(t)$ and $p^\ast_k(t)$ at time
$t=t_s$ on may then propagate the Hamiltonian equations (\ref{3.22},\ref{3.23}) up to the time
$t=t_f$. It is clear that for a typical choice of $\dx_k(t_s)$ such a ``shooting procedure''
does not lead to the desired end-result $\dx_k(t_f)=0$. But we also know from the mere existence
of the minimizing paths that there must be specific $\dx_k(t_s)$-values which do the job. 
Furthermore, eq.(\ref{4.32}) tells us that it is not necessary to proceed until $t=t_f$,
rather 
it is sufficient to take any time $t_u$ at which $x_k^\ast(t)$ has reached the $x_u(t)$-neighborhood 
and then check whether (\ref{4.32}) is satisfied.

If $t_s - t_0$ and $t_f - t_u$ are already large then the integrands in (\ref{4.31},\ref{4.32}) are
extremely small. Thus, tiny changes of $\dx_k(t_s)$ and $p^\ast_k(t_s)$ will lead to huge
changes of $t_0$ and $t_f$. Especially, by letting $t_0\to -\infty$ and $t_f\to\infty$ those
integrals vanish and one recovers the master path $x^\ast_\opt (x+k\T)$ associated
with $x_k^\ast(t)$. 
This confirms our conclusion from the previous subsection 
that a meaningful limit
of $x^\ast_k(t)$ for $t_0\to\-\infty$ and $t_f\to\infty$ exists
and that for finite but large $t_f - t_0$ the difference 
between $x^\ast_k(t)$ and the associated 
master path $x^\ast_\opt(t+k\T)$ is extremely small for all $t\in[t_0,t_f]$.

An example for which $t_f-t_u$ is {\em not} large 
is the path $x_0^\ast(t)$, {\em i.e.}, the one which crosses over from the
neighborhood of $x_s(t)$ into that of $x_u(t)$ at the latest possible moment, see fig.3.
For this path $x_0^\ast(t)$, the time $t_u$ at which it enters the neighborhood
of $x_u(t)$ is already rather close to $t_f$ an so the integral in (\ref{4.32}) 
is not any more small. As a consequence, the deviation of $x^\ast_0(t)$ from $x^\ast_\opt(t)$ 
is no longer small as $t$ approaches $t_f$. In particular, for $t=t_f$ it follows that
$x_0^\ast(t_f)-x^\ast_\opt(t_f)=-\dx_\opt(t_f)$ is no longer small and with 
(\ref{4.28}) we 
conclude that the same is true for the momentum $p^\ast_\opt(t_f)$, i.e.
\begin{equation}
%p^\ast_0(t_f)\, ,\ 
p^\ast_{{\rm opt}}(t_f)\qquad\mbox{not small}\ .
\label{4.32''}
\end{equation}

With increasing $k$-values, the deviations $-\dx_{\opt}(t+k\T)$ between 
$x_k^\ast(t)$ and the associated master path
$x^\ast_\opt(t+k\T)$ in the vicinity of $t_f$
are rapidly decreasing, essentially like $\exp\{-\lambda_{u}\, k\,\T\}$, see
(\ref{4.17},\ref{4.29}).
In the same way, for the largest possible $k$-values, $k\simeq K(t_f,t_0)$ (see (\ref{4.4'})),
corresponding to paths $x^\ast_k(t)$ with only a very short initial time segment close to $x_s(t)$,
the deviations from $x^\ast_\opt(t+k\,\T)$ are no longer small for $t$ close to $t_0$. As we 
will see later, paths $x^\ast_k(t)$ with such large $k$-values are negligible
in the sum (\ref{4.1}). For this reason, we will henceforth neglect
deviations between $x^\ast_k(t)$ and $x^\ast_\opt(t+k\,\T)$ 
and between $p^\ast_k(t)$ and $p^\ast_\opt(t+k\,\T)$ for times $t$ near 
the starting point $t_0$.
Formally, this approximation is equivalent to letting
\begin{equation}
t_0\to -\infty\ .
\label{4.32.3}
\end{equation}

\subsection{Approximations in terms of the master path}
Our next objective is to express the action (\ref{3.24}) of
the path $x^\ast_k(t)$ in terms of the associated master path
$x^\ast_\opt(t+k\T)$. We recall that while $x^\ast_k(t)$ satisfies the
boundary conditions (\ref{4.3}), those of $x^\ast_\opt(t)$
are $x^\ast_\opt(t)-x_s(t)\to 0$ for $t\to-\infty$ and 
$x^\ast_\opt(t)-x_u(t)\to 0$ for $t\to \infty$. We now modify the latter boundary condition
and require instead that
\begin{equation}
t_k:=t_f+k\, \T
\label{4.32'}
\end{equation}
is the final time and $x_k:=x^\ast_\opt(t_k)$ the final position. In other words,
we simply truncate the master path $x^\ast_\opt(t)$ at the time $t_k$, associated
with the final time $t_f$ of $x^\ast_k(t)$.
Since this ``new'' path $x^\ast_\opt(t)$ with $t\in[-\infty,t_k]$ obviously 
still satisfies the
Hamiltonian equations (\ref{3.22},\ref{3.23}) it is again an extremizing path. The
value of the action for this path follows like in (\ref{3.24}) as
\begin{equation}
\phi_\opt (x_k,t_k):=\int_{-\infty}^{t_k}p^\ast_\opt(t)^2\, dt
\label{4.32.b}
\end{equation}
and the relations (\ref{3.25},\ref{3.31}) take the form
\begin{eqnarray}
\frac{\partial\,\phi_\opt(x_k,t_k)}{\partial\, x_k} & = & p^\ast_\opt(t_k)\label{4.33}\\
\frac{\partial^2\,\phi_\opt(x_k,t_k)}{\partial\, x_k^2} & = & g^\ast_\opt(t_k)\ .\label{4.34}
\end{eqnarray}
With (\ref{3.24},\ref{4.4}) we can rewrite (\ref{4.32.b}) as
\begin{equation}
\phi_\opt (x_k,t_k)=\phi_\opt - \int_{t_k}^\infty p^\ast_\opt(t)^2\, dt\ .
\label{4.35}
\end{equation}

Next we express the action (\ref{3.17}) of the path $x^\ast_k(t)$ by expanding the 
one belonging to the associated master path $x^\ast_\opt(t+k\,\T)$ in powers
of the difference $-\dx_\opt(t_k)$ between the end-points $x^\ast_k(t_f) =
x_{u}(t_{f})$ and $x^\ast_\opt(t_f+k\,\T) = x^{\ast}_{\opt}(t_{k})$:
\begin{eqnarray}
\phi_k(x^{\ast}_k(t_{f}),t_f) & = & \phi_\opt(x_k,t_k)-
\dx_\opt(t_k)\, \frac{\partial\,\phi_\opt(x_k,t_k)}{\partial\, x_k}
\nonumber \\ 
& + & \frac{\dx_\opt(t_k)^2}{2}\, \frac{\partial^2\,\phi_\opt(x_k,t_k)}{\partial\, x_k^2} 
+ ...
\label{4.36a}
\end{eqnarray}
As justified above eq.({\ref{4.32.3}), the analogous contribution
in powers of $\dx_\opt(t_0+k\T)$ is negligible on the right hand side of
(\ref{4.36a}).
By exploiting (\ref{4.33}-\ref{4.35}) and the counterparts of (\ref{4.28}-\ref{4.30})
for $x^\ast_\opt(t+k\,\T)$, one arrives after a short calculation at
\begin{eqnarray}
\phi_k(x^{\ast}_k(t_{f}),t_f) &  = & 
\phi_\opt  + \int_{t_k}^\infty p^\ast_k(t)^2\, dt\nonumber \\
& \times & [1 + g^\ast_\opt(t_k)\, \A_u(t_k)+...\, ] \ .
\label{4.36}
\end{eqnarray}
A similar expansion of $p^\ast_\opt(t_k)$ from (\ref{4.33}) yields for $p^\ast_k(t_f)$ the
approximation
\begin{eqnarray}
p^\ast_k(t_f) & = & p^\ast_\opt(t_k) 
+ \dx_\opt(t_k)\, \frac{\partial^2\,\phi_\opt(x_k,t_k)}{\partial\, x_k^2} + ...\nonumber\\
& = & p^\ast_\opt(t_k)\, [1 + g^\ast_\opt(t_k)\, \A_u(t_k)+...\, ]\ . \label{4.38}
\end{eqnarray}

We now turn to the prefactor-terms $Q^\ast_k(t)$ in (\ref{4.1}). Within the neighborhoods of
$x_{s,u}(t)$ for which the approximations (\ref{4.5})
and thus (\ref{4.6}) are valid, we can infer from (\ref{3.27}) that
\begin{equation}
\dot Q^\ast_k(t)/2 - Q^\ast_k(t)\, F'(x_{s,u}(t),t) = \mbox{const.} =: \mu_{s,u} \ .
\label{4.39}
\end{equation}
By comparison with (\ref{3.29}) we further see that
\begin{equation}
g^\ast_k(t)\, Q^\ast_k(t) = \mu_{s,u} \ .
\label{4.40}
\end{equation}

The constant $\mu_s$, which is connected with the neighborhood of $x_s(t)$ and is
typically different form $\mu_u$, follows from the initial conditions (\ref{3.28})
as $\mu_s = 1/2$. Hence, the solution of (\ref{4.39}) takes the form
\begin{equation}
Q^\ast_k(t) = \A_s(t,t_0)/2 \ .
\label{4.41}
\end{equation}
As a by-product we find from (\ref{3.31}), evaluated for an arbitrary final 
condition $t_f=t$ and $x_f=x$ in combination with (\ref{4.40},\ref{4.41}) that
\begin{eqnarray}
\frac{\partial^2 \phi _k(x,t)}{\partial x^2} & = & \frac{1}{\A_s(t,t_0)} \ .
\label{4.41'}
\end{eqnarray}

Within the linearization (\ref{4.5}), closer inspection shows that
only a single summand appears in the conditional probability
(\ref{3.26}) and one recovers the expected Gaussian result
for $x$ close to the stable periodic orbit $x_s(t)$:
\begin{eqnarray}
& & p(x,t\, | \, x_s(t_0), t_0) = \nonumber \\
& &\left(\frac{1}{2\,\pi\, D\, \A_{s}(t,t_{0})}\right)^{\frac{1}{2}}\ 
\exp\left\{-\frac{[x-x_s(t)]^2}{2\, D\, \A_{s}(t,t_{0})}\right\}\ .
\label{4.42}
\end{eqnarray}

Returning to (\ref{4.41}), it is remarkable that besides the initial time $t_0$ no further 
details of the path $x^\ast_k(t)$ play a role. Especially, if $x^\ast_k(t)$ remains for a long time
in the neighborhood of $x_s(t)$ where (\ref{4.41}) is valid, then by the time it leaves 
this neighborhood, say $t=t_s$, the quantity $\A_s(t,t_0)$ is practically equal to $\A_s(t)$ 
from (\ref{4.21}) and thus $Q^\ast_k(t)$ equal to the associated master-prefactor 
$Q^\ast_\opt(t+k\,\T)$. Within our usual approximation (\ref{4.32.3}) we thus have
\begin{eqnarray}
Q_k^\ast (t_s) & = & Q_\opt^\ast (t_s+k\T ) = \A_s(t_s)/2\label{4.42'}\\
\dot Q_k^\ast (t_s) & = & \dot Q_\opt^\ast (t_s+k\T ) = \dot \A_s(t_s)/2 \ .\label{4.42''}
\end{eqnarray}
These relations are then used as initial conditions in (\ref{3.27})
in order to propagate  $Q_k^\ast (t_s)$ and $Q_\opt^\ast (t_s+k\T )$ through the 
``crossover-segments'' of the corresponding paths
$x_k^\ast (t_s)$ and $x_\opt^\ast (t_s+k\T )$
up to a certain time-point, say $t=t_u$,
beyond which the linearization (\ref{4.5}) about $x_u(t)$ and thus (\ref{4.39})
can be applied. 

Once the neighborhood of $x_u(t)$ is reached, {\em i.e.}, for $t\geq t_u$,
the solution of (\ref{4.39},\ref{4.40}) can be written with (\ref{4.13}) as
\begin{equation}
Q^\ast_k(t) = Q^\ast_k(t_u)\, e^{2\,\J_u(t,t_u)} \, [ 1-g^\ast_k(t_u)\, \A_u(t_u,t)]\ .
\label{4.45.1}
\end{equation}
In view of (\ref{4.10}) this yields furthermore
\begin{eqnarray}
Q^\ast_\opt(t)\, p^\ast_\opt(t)^2 & = & Q^\ast_\opt(t_u)\, p^\ast_\opt(t_u)^2
\nonumber \\  
& \times & [ 1-g^\ast_\opt(t_u)\, \A_u(t_u,t)]\ .
\label{4.45.2}
\end{eqnarray}
Due to (\ref{4.22}), the factor $\A_u(t_u,t)$ approaches $-\A_u(t_u)$ as $t-t_u$ becomes large.
It follows that the left hand side of (\ref{4.45.2}) tends towards a
finite limit as $t\to\infty$:
\begin{equation}
q_\opt := \lim_{t\to\infty} Q^\ast_\opt(t)\, p^\ast_\opt(t)^2\ .
\label{4.47}
\end{equation}
Since $t_u$ is an arbitrary reference-time in (\ref{4.45.2}), we can first let $t\to\infty$ and 
then rename $t_u$ as $t$ with the result
\begin{equation}
Q^\ast_\opt(t) = \frac{q_\opt}{p^\ast_\opt(t)^2}-\mu_\opt\, \A_u(t)\ ,
\label{4.48}
\end{equation}
where the (finite) constant $\mu_\opt$ is defined analogously to (\ref{4.40}) as
\begin{equation}
\mu_\opt := \lim_{t\to\infty}g^\ast_\opt(t)\, Q^\ast_\opt(t)\ .
\label{4.49}
\end{equation}
Exploiting once more (\ref{4.40}) and (\ref{4.49}), we can eliminate $Q^\ast_\opt(t)$ in 
(\ref{4.48}) in favor of $g^\ast_\opt(t)$ with the result
\begin{equation}
g^\ast_\opt(t)=\frac{p^\ast_\opt(t)^2}{q_\opt/\mu_\opt -p^\ast_\opt(t)^2 \A_u(t)} \ .
\label{4.49'}
\end{equation}

As discussed below (\ref{4.32}), the deviations of $x^\ast_k(t)$ from the 
associated
master path $x^\ast_\opt(t+k\T)$ become smaller and smaller as $k$ increases and
in view of (\ref{4.42'},\ref{4.42''},\ref{3.27}) we expect a similar convergence of $Q^\ast_k(t)$
towards $Q^\ast_\opt(t+k\T)$. In Appendix C, the following quantitative estimate 
for this convergence is established
for all times $t\in[t_u,t_f]$:
\begin{equation}
Q^\ast_k(t) = Q^\ast_\opt(t+k\,\T)\, [1+\Ord (p^\ast_\opt(t_k)^2)]\ ,
\label{4.45'}
\end{equation}
where the order of magnitude is meant with respect to the dependence
on $k$.

 From the technical viewpoint, 
Eq.(\ref{4.48}) in combination with (\ref{4.45'}) is a central 
and highly non-trivial result of our present work.
Since $\A_u(t)$ is periodic in $t$ and since
$p^\ast_\opt(t)$ decreases exponentially according to
(\ref{4.10}), we see from (\ref{4.48},\ref{4.45'}) 
that the prefactor $Q^\ast_k(t)$ diverges exponentially with the
time which the path $x^\ast_k(t)$ spends in the neighborhood of $x_u(t)$, in striking contrast
to the behavior (\ref{4.41}) close to $x_s(t)$. The basic physical reason for 
this divergence of
$Q^\ast_k(t)$ is that the probability of a stochastic process (\ref{2.3})
to permanently remain close to the unstable periodic orbit $x_u(t)$ decreases exponentially
with increasing time.
Since typically the process closely follows a deterministic trajectory,
the action barely grows and thus it is  the prefactor $1/Q^\ast_k(t_f)^{1/2}$ in
(\ref{3.26}) which has to account for the exponential decrease in time!

Since $p^\ast_\opt(t_k)$ decreases exponentially with $k$ we see from
(\ref{4.49'}) that $g^\ast_\opt(t_k)$ tends to zero like 
$p^\ast_\opt(t_k)^2$. In view of (\ref{4.33},\ref{4.34}) we therefore
conjecture that also higher derivatives of $\phi_\opt(x_k, t_k)$ continue
to scale like the corresponding powers of $p^\ast_\opt(t_k)$. The terms
indicated by the dots in (\ref{4.36a}-\ref{4.38}) are
then indeed negligibly small.

\subsection{Evaluation and discussion of the rate}
We are now in the position to evaluate the rate-formula (\ref{4.1})
in terms of the master path $x^\ast_\opt(t)$. To this end we
approximate in (\ref{4.36})
and (\ref{4.38}) the square brackets by $1$, neglect in 
(\ref{4.45'}) the term of order $p^\ast_\opt(t_k)^2$
and in (\ref{4.48}) the last term (being also a correction 
of order $p^\ast_\opt(t_k)^2$).
By dropping the index of $t_f$ we then can infer from (\ref{4.1},\ref{4.4'}) 
our central result for the {\em instantaneous rate} \cite{leh00} 
\begin{equation}
\Gamma (t) = \sqrt{D}\,\alpha_\opt\, e^{-\phi_\opt / D}\,\kappa_\opt(t,D)
%\, [1+E(D)]
\, [1+\Ord (D^\gamma)]
\label{4.51}
\end{equation}
%\alpha_\opt & := & \frac{1}{\T \sqrt{4\pi}}\,
%\lim_{t\to\infty}\frac{1}{\sqrt{p^\ast_\opt(t)^2\, Q^\ast_\opt(t)}}\label{4.52}\\
\begin{equation}
\alpha_\opt   :=  [4\, \pi\,  \T^2
\lim_{t\to\infty} p^\ast_\opt(t)^2\, Q^\ast_\opt(t) ]^{-1/2} 
\label{4.52}
\end{equation}
\begin{eqnarray}
\kappa_\opt(t,D) & := & \T\sum\limits_{k=0}^{K(t,t_0)}
\frac{p^\ast_\opt(t+k\T)^2}{D}\nonumber\\
& \times &
\exp\left\{-\frac{1}{D}\int_t^\infty p^\ast_\opt(t'+k\T)^2\, dt'\right\} \ .
\label{4.53}
\end{eqnarray}

The effect of our various approximations in deriving this result
together with the corresponding 
%relative error $E(D)$
``accuracy exponent'' $\gamma >0$
in (\ref{4.51}) will be discussed in Sect.~IV.E below. 
Next, we analyze in more detail the properties of $\kappa_\opt(t,D)$.
By means of (\ref{4.10},\ref{4.17},\ref{4.23}) we rewrite (\ref{4.53}) as
\begin{eqnarray}
\kappa_\opt(t,D) & = & \T\sum\limits_{k=0}^{K(t,t_0)}
\frac{p^\ast_\opt(t)^2\, C^k}{D}\nonumber\\
& \times &
\exp\left\{-\frac{p^\ast_\opt(t)^2\, C^k\, \A_u(t)}{2\,D}\right\} \label{4.53'}\\
C & := & e^{-2\,\lambda_u\, \T} \ . \label{4.54}
\end{eqnarray}
Since $0<C<1$ there is a competition in the sum
(\ref{4.53'}) between the exponential terms which increase with $k$ and the
pre-exponential factors which decrease with $k$. One readily sees that the
dominant contribution to the sum stems from a few $k$-values around the
real number $\hat k$, implicitly defined via
\begin{equation}
p^\ast_\opt(t)^2\, C^{\hat k}\, \A_u(t) = 2\,D\ .
\label{4.55}
\end{equation}
Recalling that $t$ stands here for $t_f$ and
since neither $\A_u(t=t_f)$ nor $p_\opt^\ast(t=t_f)$ 
(cf. (\ref{4.23},\ref{4.32''})) are small
quantities, it follows that
$\hat k$ is, for small noise-strengths $D$, much larger
than $0$ but, for sufficiently large $t_f-t_0$, according to 
(\ref{4.4'}) also much smaller than $K(t=t_f,t_0)$. Therefore the sum in
(\ref{4.53'}) and thus in (\ref{4.53}) can be extended to arbitrary integers
$k$ at the price of an error which is {\em exponentially} small in $D$,
%As we will see below, this error is negligibly small in comparison with
%the relative error $E(D)$ which is already present 
{\em i.e.}, without actually affecting the accuracy exponent $\gamma$
in (\ref{4.51}).
As a further consequence of the fact that the dominant $k$-values
are much smaller than the upper limit $K(t,t_0)$ for large $t-t_0$, we see 
that our formal approximation (\ref{4.32.3}) is indeed self-consistently 
satisfied.

Next we notice that under the sum in (\ref{4.53}), the pre-exponential
term is nothing else than the time-derivative of the expression in the 
exponential. By extending the sum over all integer $k$-values as
justified above we obtain
\begin{eqnarray}
& & \frac{1}{\T}\int_t^{t+\T}\kappa_\opt(t',D)\, dt' \nonumber\\
& & = \sum_{k=-\infty}^{\infty} \exp\left\{-\frac{1}{D}\int_{t}^{\infty}
  p^{\ast}_{\opt}(t')^{2} dt'\right\} \bigg|_{t=k\,\T}^{(k+1)\T}\nonumber\\
& & = 1 - \exp\left\{-\frac{1}{D}\int_{-\infty}^{\infty} 
              p^{\ast}_{\opt}(t')^{2} dt'\right\} \ .
\label{4.57'}
\end{eqnarray}
Neglecting as usual errors exponentially small in $D$
this leads us to the remarkable conclusion that
\begin{eqnarray}
& & \frac{1}{\T}\int_t^{t+\T}\kappa_\opt(t',D)\, dt' = 1
\label{4.57}
\end{eqnarray}
for all $t$ and all (small) $D$. For the {\em time averaged rate} (\ref{2.21}) we 
thus obtain from (\ref{4.51},\ref{4.57}) our central result \cite{leh00}
\begin{eqnarray}
\bar \Gamma = \sqrt{D}\,\alpha_\opt\, e^{-\phi_\opt / D}
%\, [1+E(D)] \ .
\, [1+\Ord (D^\gamma)] \ .
\label{4.60}
\end{eqnarray}
It consists of an Arrhenius-type exponentially leading part with an
``effective potential barrier'' $\phi_\opt$ and a non-trivial
pre-exponential $D$-dependence.
The two quantities $\alpha_\opt$ and $\phi_\opt$
follow from the master-path $x^\ast_\opt(t)$ according to (\ref{4.4})
and (\ref{4.52}). Thus they are independent of $D$ but
depend in a highly nontrivial way
on various global properties of the deterministic force field $F(x,t)$
in (\ref{2.3}). In general, their explicit value can only be determined
numerically or by means of approximations.
An exactly analytically solvable special case will be presented in Sect.~V.A.

We recall that for equilibrium systems, characterized by
a time-independent force field $F(x)= - V'(x)$ in (\ref{2.3}),
the escape rate exhibits an exponentially leading 
Arrhenius factor, which involves simply the barrier 
against escape of the static potential
$V(x)$, and a $D$-independent
pre-exponential factor which depends only on local properties of 
the potential at the barrier and the well \cite{han90},
see also (\ref{5.27}) below.
The different structure of (\ref{4.60}) is thus a consequence of the
far from equilibrium situation created by the time-dependence of the
deterministic force field $F(x,t)$.

As announced in Sect.~II.C, the time-averaged escape rate
for the periodic force field $F(x,t)$ can be identified with
that of its supersymmetric partner force 
field (\ref{susy1}) for asymptotically weak noise $D$
without any further restrictions on $F(x,t)$. 
The detailed proof of this highly non-trivial
invariance property
of (\ref{4.60}) is carried out in Appendix D.

Returning to the instantaneous rate (\ref{4.51}), we see that
it exhibits in comparison with
the time averaged rate (\ref{4.60}) the additional time-dependent 
factor $\kappa_\opt(t,D)$. The explicit evaluation 
of this factor requires the knowledge of
one more global quantity, for instance of
\begin{equation}
\beta_\opt(t) := \lim_{\hat{t}\to\infty}p^\ast_\opt(\hat{t})\, e^{\J_u(\hat{t},t)}.
\label{4.60'}
\end{equation}
Note that due to relation (\ref{4.16}) the $t$-dependency of this quantity is
actually quite simple.
According to (\ref{4.10}), this definition (\ref{4.60'}) allows us to rewrite
(\ref{4.53'}) -- with the range of $k$ extended to arbitrary integers -- as
\begin{eqnarray}
\kappa_\opt(t,D) & = & \T\sum\limits_{k=-\infty}^{\infty}
\frac{\beta_\opt(t)^2 \, C^k}{D}\nonumber\\
& \times &
\exp\left\{-\frac{\beta_\opt(t)^2 \, C^k\, \A_u(t)}{2\,D}\right\}\ . 
\label{4.63}
\end{eqnarray}
Besides $\beta_\opt(t)$ all other quantities in this expression are 
determined by local 
properties of the force field $F(x_u(t),t)$ along the unstable periodic orbit.
By exploiting (\ref{4.17},\ref{4.24},\ref{4.54}) it follows that
\begin{eqnarray}
\kappa_\opt(t+\T,D) & = & \kappa_\opt(t,D) \label{4.58}\\
\kappa_\opt(t,C\,D) & = & \kappa_\opt(t,D) \label{4.59}\ .
\end{eqnarray}
Together with (\ref{4.55}) and the obvious property $0<\kappa_\opt(t,D)<\infty$
this completes our qualitative picture of the way in which
$\Gamma (t)$ oscillates around its average value $\bar\Gamma$.

\subsection{The accuracy exponent $\gamma$}
In the following, we come to the determination
of the accuracy exponent $\gamma$
in (\ref{4.51},\ref{4.60}). We will not elaborate here all the
details of the rather involved calculations but restrict ourselves 
to the main steps. 

First of all, we recall that a contribution $\Ord (D)$ is
inherited right away from formula (\ref{4.1}). Next we have approximated
the square brackets in (\ref{4.36}) by $1$.
For those $k$-values which mainly contribute to the rate it can be inferred
from (\ref{4.55}) together with
$t=t_f$ and (\ref{4.32'}) that
\begin{equation}
p^\ast_\opt(t_k)^2 = \Ord (D)
\label{4.61}
\end{equation}
and hence with (\ref{4.49'}) that
\begin{equation}
g^\ast_\opt(t_k) \, \A_u(t_k) = \Ord (D)\ .
\label{4.62}
\end{equation}
Since the integral in (\ref{4.36}) is of the same order of magnitude as
$p^\ast_\opt(t_k) $ from (\ref{4.61}) we conclude that the total error 
we committed in (\ref{4.36}) is of the order $\Ord (D^2)$, thus contributing 
once more a term of the order $\Ord (D)$ in the rate-formulas (\ref{4.51},\ref{4.60}).
The same conclusion can be drawn with respect to our
approximating the square brackets by $1$ in (\ref{4.38})
and neglecting the $\Ord (p^\ast_\opt(t_k) ^2)$ -term in (\ref{4.45'})
as well as the last term in (\ref{4.48}).
In other words, the relative error induced by all our so far
made approximations is of the order $\Ord (D)$.

What remains is a closer
inspection of the approximation (\ref{4.5}) for $F(x,t)$ in the neighborhood
of $x_{s,u}(t)$.
One readily sees that actually only the approximation in the neighborhood
of the unstable periodic orbit 
$x_u(t)$ matters in our quantitative evaluation of the rate; the
basic reason for this is once more our assumption $t_{0}\to-\infty$ in~(\ref{4.32.3}).
In case that (\ref{4.5}) happens to be an exact identity in this neighborhood
of $x_u(t)$,
then the total error committed in the rate-formulas (\ref{4.51},\ref{4.60}) is thus
of the order $\Ord (D)$. Otherwise, a closer analysis of the relevant
perturbative corrections shows that the error introduced via the approximation
(\ref{4.5}) amounts to corrections of the order $\Ord (p^\ast_\opt(t_k) )$ in
the rate-formula, {\em i.e.}, of the order $\Ord (\sqrt{D})$ according to
(\ref{4.61}). In other words, we can conclude that
\begin{eqnarray}
& & \gamma = 
\left \{ \begin{array}{ll} 
1 & \mbox{if}\ \ F''(x_u(t),t)\equiv 0\\
1/2 & \mbox{otherwise.}
\end{array} \right .
\label{4.64}
\end{eqnarray}

In the case $\gamma =1/2$ it is important that in the global quantities 
$\phi_\opt$, $\alpha_\opt$, $\beta_\opt(t)$ from (\ref{4.4},\ref{4.52},\ref{4.60'})
the long-time limits are made and the exact master path is utilized without
any further approximations. If instead in these definitions any finite
reference-time in combination with relations based on the approximation
(\ref{4.5}) were used, then this would introduce a possibly very small
but nevertheless $D$-independent error and so ruin the asymptotically
exact predictions (\ref{4.51},\ref{4.60}) in the weak noise limit $D\to 0$.

In cases for which (\ref{4.5}) is not exactly satisfied in the
neighborhood of the unstable periodic orbit $x_u(t)$ and
hence $\gamma = 1/2$, it is in principle possible to calculate
perturbatively the corresponding corrections such as to arrive again 
at a reduced relative error $\Ord (D)$ in the so improved rate-formulas,
though the actual calculations and the resulting expressions
become very complicated. On the other hand, further reducing the
$\Ord (D)$-error is even in principle rather problematic
since it would require to go beyond the saddle point
approximation in the path-integral approach from Sect.~III.

At this point it may also be worth to recall from Sect.~IV.A that 
for any fixed (however small) $D$-value, the
error $\Ord (D)$ in (\ref{4.1}), which is inherited by the final rate-formula, 
diverges as the amplitude of the time 
dependency of $F(x,t)$ tends to zero, but also if its period $\T$
either tends to zero or to infinity. Thus, neither of these
limits commutes with the limit $D\to 0$.

\subsection{The limits $t\to\infty$ and $D\to 0$}
In the derivation of the rate-formula (\ref{4.51}) we have assumed that all
paths $x^\ast_k(t)$ which notably contribute in (\ref{4.1}) sojourn for a 
very long initial time-interval close to the stable periodic orbit $x_s(t)$,
see (\ref{4.32.3}). On the other hand, eq. (\ref{4.55}) tells us that the 
amount of time which those dominant paths spend in the neighborhood of the
unstable periodic orbit $x_u(t)$ is roughly speaking of the order 
$\Ord (\ln 1/D)$. Both these conditions are compatible only if $t-t_0$ 
substantially exceeds in order of magnitude $\ln 1/D$. In the physically 
relevant case, the noise-strength $D$ is small but finite and this
condition is well satisfied after a comparatively short 
``transient'' time-period. Thus, strictly speaking, in (\ref{4.51}) and
(\ref{4.60}), with decreasing $D$-values also the lower limit of the admitted 
times $t-t_0$ is tacitly assumed to slowly increase in proportion
to $\ln 1/D$.

We remark that our result (\ref{4.51})
obviously remains periodic in $t$ for arbitrarily large $t-t_0$, see
(\ref{4.58}). Therefore, the restriction of the utilized basic formula
(\ref{2.21}) to values of $t-t_0$ much smaller than $1/\bar\Gamma$
does no longer apply to the final result (\ref{4.51}), see
also the discussion below (\ref{2.22}).

In the physically less relevant case that $t-t_0$ is kept at an
arbitrary but fixed value and then $D$ is made smaller and smaller, 
things become different. As pointed out in Sect.~IV.A, for any finite 
initial and final times $t_0$ and $t=t_f$, there exists generically a 
unique absolute minimum $x^\ast_{k_0}(t)$ of the action.
For sufficiently small $D$ the $k_0$ -term will thus completely 
dominate the sum in (\ref{4.1}), {\em i.e.},
\begin{equation}
\Gamma (t=t_f) = 
\frac{p^\ast_{k_0} (t_f)\, 
e^{- \phi_{k_0} (x_u(t_f), t_f)/D}}
{[4\,\pi\, D\, Q^\ast_{k_0}(t_f)]^{1/2}} \ .
\label{4.65}
\end{equation}
While most of the quantities on the right hand side of this result 
(including the index $k_0$) still
depend in a very complicated way on the time $t=t_f$, no additional
implicit $D$-dependence is hidden. The most striking feature is the
$1/\sqrt{D}$ pre-exponential behavior in comparison with the
$\sqrt{D}$ -scaling in (\ref{4.51})!

Qualitatively, the crossover from (\ref{4.65}) to
(\ref{4.51}), either as $t$ increases or as $D$ decreases, is clear:
At some point the $k$-dependence of the pre-exponential factors in (\ref{4.1})
is no longer negligible in comparison
with the exponentially leading contributions and so the
dominant $k$-value moves away from $k_0 =k_0(t,t_0)$ towards smaller
values $k\simeq \hat k$, cf. (\ref{4.55}). At the same time, more that one
term in the sum (\ref{4.1}) starts to notably contribute.

Quantitatively, a leading order approximation follows
along the same line of reasoning as in the derivation of (\ref{4.51}) from
(\ref{4.1}), except that in the approximation for the action 
(\ref{4.36a}), also contributions due to the deviations between $x^\ast_k(t)$
and its associated master-path $x^\ast_\opt(t+k\T)$ at times close to $t_0$
have to be included, that is, the approximation (\ref{4.32.3}) should be
abandoned. The final result is again of the same form as in (\ref{4.51}) but
with a larger error than $\Ord (D^\gamma)$ and instead of (\ref{4.53}) with
\begin{eqnarray}
& & \kappa_\opt(t,D) := T\sum\limits_{k=0}^{K(t,t_0)}
\frac{p^\ast_\opt(t+k\T)^2}{D}\nonumber\\
& & \times
\exp\left\{-\frac{1}{D}\left[\int_t^\infty + \int_{-\infty}^{t_0}\right]
p^\ast_\opt(t'+k\T)^2\, dt'\right\} \ .
\label{4.66}
\end{eqnarray}
For moderate $t-t_0$ or extremely small $D$ the exponential in (\ref{4.66}) 
depends very strongly on $k$ and therefore the sum is dominated
by a single term $k=k_0(t,t_0)$. Upon increasing $t-t_0$ or $D$
this strong $k$-dependence of the exponential and hence the dominance
of the $k_0$-contribution is softened and the already discussed
qualitative crossover-behavior is recovered.

\subsection{More general seeds $x_0$}
So far, our rate-formula (\ref{4.51}) is restricted to the case (\ref{4.2}) 
that the initial position $x_0$ at time $t_0$ coincides with the stable
periodic orbit $x_s(t_0)$. As pointed out in Sect.~II, one expects that for 
large enough times $t-t_0$ the initial position $x_0$ should not matter,
provided it is chosen inside the domain of attraction of $x_s(t)$.
For sufficiently small noise-strengths $D$ this is the case whenever
\begin{equation}
x_0 < x_u(t_0)\ .
\label{4.67}
\end{equation}
In the following, we analyze this intuitive expectation in some more quantitative
detail.

For arbitrary but fixed $x_0$ satisfying (\ref{4.67}), the observation from 
Sect.~IV.A remains true, namely that only minimizing
paths $x^\ast_k(t)$ play a role in the rate (\ref{4.1}) which closely follow 
a deterministic behavior $\dot x_k^\ast(t)\simeq F(x^\ast_k(t),t)$ for most
of the time. This requirement can be fulfilled in two basic ways and appropriate
compromises thereof. The first possibility is that
the path $x^\ast_k(t)$ closely
approximates a deterministic trajectory for a very long initial time-interval.
During this time, $x^\ast_k(t)$ approaches the periodic attractor
$x_s(t)$ very closely and does 
practically not accumulate any action (\ref{3.17},\ref{3.18}). 
Consequently, one is back to the case (\ref{4.2}) 
after an appropriate redefinition of the initial time $t_0$.
Regarding the prefactor $Q^\ast_k(t)$, one can, according to (\ref{3.23}),
approximate $p^\ast_k(t)$ in (\ref{3.27}) by zero. With the initial
conditions (\ref{3.28}) one then recovers the same solution as in
(\ref{4.41}) except that in (\ref{4.12},\ref{4.13}) the function
$\J_s(t,t_0)$ is now defined as $\int_{t_0}^t F'(x_k^\ast(t'),t')\, dt'$.
Since $x^\ast_k(t)$ practically agrees with $x_s(t)$
during a very long time interval, one sees that also with respect
to the prefactor $Q^\ast_k(t)$ we are back to the case (\ref{4.2}).
As before, we may label such paths by low $k$-values and their
contributions to the rate (\ref{4.1}) are identical to those
of the low $k$-values in (\ref{4.51}-\ref{4.53}).

The second possibility is that the minimizing path $x^\ast_k(t)$
travels from its starting point $x_0$ immediately into
the neighborhood of the unstable periodic orbit $x_u(t)$ and then very
closely follows this deterministic trajectory $x_u(t)$ for the rest
of its time. If $x_0$ is already close to $x_u(t_0)$, such paths lead to a very
small value of the action in (\ref{3.17},\ref{3.18}) and thus
will ultimately dominate the rate (\ref{4.1}) if $t-t_0$ is kept
fixed and $D$ becomes asymptotically small. 
This puzzling observation has lead to some amount of confusion in the
recent literature \cite{man99,vug99}.
The resolution is that, much like in Sect.~IV.F,
things become very different for a small but fixed $D$ in
combination with larger and larger times $t-t_0$.
The salient point is that the price to be paid for a long sojourn close to
$x_u(t)$ is a very small prefactor $p^\ast_k(t_f)/[Q^\ast_k(t_f)]^{1/2}$ in
(\ref{4.1}), as discussed below (\ref{4.45'}), namely of the order
$\exp\{-2\,\lambda_u\cdot (t-t_0)\}$. As a consequence, the
paths with low $k$-values, as discussed in the
preceding paragraph, will dominate in spite of their unfavorable 
action. Therefore, the rate-formula (\ref{4.51})
applies for any $x_0$ satisfying (\ref{4.67}) on condition that
\begin{equation}
t-t_0\gg \phi_\opt / (2 D \lambda_u)\ .
\label{4.68}
\end{equation}
This condition characterizes the asymptotic time regime for which the
rate-formula (\ref{4.51}) is valid in the case of a general initial
condition. Even for rather small $D$, the preceding transient regime
is typically confined to a few driving periods $\T$, as
illustrated by the examples in Sect.~V.
Note that (\ref{4.68}) comprises the condition from
Sect.~IV.F that $t-t_0$ has to substantially exceed in order of magnitude
$\ln 1/D$. In other words, for a generic initial condition,
eq. (\ref{4.68}) is the only restriction for the
rate-formula (\ref{4.51}), apart from the exclusion of vanishing driving
amplitudes and vanishing or diverging periods $\T$.

\subsection{Summary from the practical viewpoint}
Given an arbitrary time-periodic force field $F(x,t)$ that satisfies the 
condition in~(\ref{2.14c}), what are the necessary practical
(numerical or analytical) steps for an explicit
quantitative evaluation of the rate (\ref{4.51})?

The first step is the determination of the stable and unstable periodic 
orbits $x_{s}(t)$ and $x_{u}(t)$. An efficient way to do this is by
evolving the deterministic dynamics forward and backward over a
long time,
respectively, with a reasonably well chosen initial condition.
Once $x_{s,u}(t)$ is known, the functions $\J_{s,u}(t,t_1)$ from
(\ref{4.12}) and $\A_{s,u}(t,t_1)$ from (\ref{4.20},\ref{4.22})
follow readily, with $t_1$ being an arbitrary reference time.

The next step is the determination of the master path $x^\ast_\opt(t)$.
To this end, we chose an arbitrary but fixed time $t_s$ and a very small 
but finite positive number $\Delta x_{{\rm min}}$, characterizing the
neighborhood of $x_s(t)$ within which we are willing to accept the
errors introduced by the approximation (\ref{4.5}). We now consider the
quantity $\dx_\opt(t_s)$ as a parameter that may take values
in the interval $[\Delta x_{{\rm min}},\, \Delta x_{{\rm min}}e^{-\lambda_s\, \T}]$.
Each such parameter value $\dx_\opt(t_s)$ yields a set of initial conditions
\begin{eqnarray}
x^\ast_\opt(t_s) & = & x_s(t_s) + \dx_\opt(t_s)\label{4.69a}\\
p^\ast_\opt(t_s) & = & \dx_\opt(t_s)/\A_s(t_s) \label{4.69b}\ ,
\end{eqnarray}
see (\ref{4.9},\ref{4.31}). With these initial conditions one then evolves
$x^\ast_\opt(t)$ and $p^\ast_\opt(t)$ according to (\ref{3.22},\ref{3.23}).
For a generic value of the parameter $\dx_\opt(t_s)$, the path 
$x^\ast_\opt(t)$ will either reach $x_u(t)$ after a finite time and then
proceed towards $x=\infty$ or never reach $x_u(t)$ and instead return
into the vicinity of $x_s(t)$ as $t$ grows. By fine-tuning $\dx_\opt(t_s)$
one has to find a path $x^\ast_\opt(t)$ which remains close to $x_u(t)$
as long as possible, say until $t=t_{\rm{max}}$.
Upon varying $\dx_\opt(t_s)$ within 
$[\Delta x_{{\rm min}},\, \Delta x_{{\rm min}}e^{-\lambda_s\, \T}]$
the existence of at least one such path is guaranteed by the theory.
A second solution, corresponding to a saddle point instead of a 
minimum of the action, is also to be expected (see Sect.~V.A, below
eq.~(\ref{5.24})). 
Further local extrema 
may coexist as well. Among them, the desired solution 
$x^\ast_\opt(t)$ is the one with the smallest value of the action
\begin{equation}
\phi_\opt = \frac{\dx_\opt(t_s)^2}{2\, \A_s(t_s)} + \int_{t_s}^{t_{\rm{max}}}
p^\ast_\opt(t)^2\, dt\ ,
\label{4.70}
\end{equation}
see (\ref{3.24},\ref{4.4},\ref{4.10},\ref{4.25}). By  approximating $\hat{t}$
in (\ref{4.60'}) by $t_{\rm max}$ we obtain
\begin{equation}
\beta_\opt(t) = p^\ast_\opt(t_{\rm{max}})\, e^{\J_u(t_{\rm max},t)}\ ,
\label{4.71}
\end{equation}
whence $\kappa_\opt(t,D)$ from (\ref{4.63}) follows with $C$ from
(\ref{4.54}). Finally, one chooses the initial conditions
\begin{equation}
Q^\ast_\opt(t_s) = \A_s(t_s)/2\ ,\ \ \dot Q^\ast_\opt(t_s) = \dot \A_s(t_s)/2\ ,
\label{4.72}
\end{equation}
see (\ref{4.42'},\ref{4.42''}), and then propagates $Q^\ast_\opt(t)$ according 
to (\ref{3.27}) until $t=t_{\rm{max}}$, to obtain
\begin{eqnarray}
\alpha_\opt =  
%1/\T \sqrt{4\pi\, p^\ast_\opt(t_{\rm{max}})^2\, Q^\ast_\opt(t_{\rm{max}}}\ ,
[ 4\, \pi\, T^2\, p^\ast_\opt(t_{\rm{max}})^2\, 
Q^\ast_\opt(t_{\rm{max}}) ]^{-1/2}\ ,
\label{4.73}
\end{eqnarray}
see (\ref{4.52}).

The accuracy of $\phi_\opt$, $\beta_\opt(t)$, $\alpha_\opt$ from
(\ref{4.70},\ref{4.71},\ref{4.73}) can be estimated by observing how
little these quantities change if $t_{\rm{max}}$ is varied and if 
$\Delta x_{\rm{min}}$ is changed by a factor $e^{\lambda_s \T}$.

We finally note that the association of $x^\ast_\opt(t+k\T)$
with the specific path $x^\ast_k(t)$ as in the preceding subsections 
does not play a role any more in the above described practical procedure.

\section{Examples}

In general, the explicit quantitative evaluation of $\phi_{\opt}$,
$\alpha_{\opt}$, and $\kappa_{\opt}(t,D)$ in the rate-formula (\ref{4.51}) is
not possible in closed analytical form. Exceptions are piecewise parabolic
potentials $V(x)$ in conjunction with an additive sinusoidal driving
(\ref{2.5}). In Subsect.~A, the simplest example \cite{leh00}
with two parabolic pieces
will be worked out and compared with accurate numerical results and with the
approximation for small driving amplitudes from \cite{sme99}. In Subsect.~B we
elaborate as a second example the case of a force field (\ref{2.5}) deriving
from a cubic potential $V(x)$ along the lines of the numerical recipe from
Sect.~IV.H.

\subsection{Piecewise parabolic potential}

We consider the force field from (\ref{2.5}) with a piecewise parabolic
potential of the form 

\begin{eqnarray}
V(x\le 0) & = & \frac{\lambda_s}{2}[\xs^2 - (x-\xs)^2]\nonumber\\
V(x\ge 0) & = & \frac{\lambda_u}{2}[\xu^2 - (x-\xu)^2],
\label{5.1}
\end{eqnarray}
where $\xs$ denotes the potential well (stable fixed point) and $\xu$ the
saddle (unstable fixed point), with the properties
\begin{equation}
\xs < 0 \ \ , \ \ \ \xu > 0\ \ .
\label{5.2}
\end{equation}
The parameters
\begin{equation}
\lambda_s < 0 \ \ , \ \ \ \lambda_u > 0
\label{5.3}
\end{equation}
characterize the piecewise constant curvatures and thus the time scales
(Lyapunov exponents) of the deterministic motion near the attractor $\xs$ and
the repeller $\xu$, respectively.  The force-field (\ref{2.5}) then takes the
explicit form 
\begin{eqnarray}
   F(x\le 0,t) &=& \lambda_s ( x - \xs ) + A\,\sin(\Omega\, t)\nonumber\\
   F(x\ge 0,t) &=& \lambda_u ( x - \xu ) + A\,\sin(\Omega\, t) \ .
\label{5.4}
\end{eqnarray}
In particular, the quantities $\lambda_{s,u}$ in (\ref{5.1},\ref{5.4}) are
identical to those from (\ref{4.17})

Requiring continuity at $x = 0$ we conclude from (\ref{5.4}) that $\lambda_s
\xs = \lambda_u\xu$.  Selecting as independent model parameters $\lambda_s$,
$\lambda_u$, and the static potential barrier
\begin{equation}
\Delta V := V(\xu )- V(\xs )
\label{5.5}
\end{equation}
the fixed points $\bar x_{s,u}$ can be expressed through
\begin{equation}
  \lambda_{s} \xs = 
  \lambda_{u} \xu =  
  \sqrt{\frac{2\,\Delta V |\lambda_{s}|\lambda_{u}}{|\lambda_{s}|+\lambda_u}}
  \label{5.6}
\end{equation}

Turning to the determination of the stable and unstable periodic orbits
(\ref{2.14a}) it is convenient to make a somewhat stronger assumption than in
(\ref{2.14c}) namely that both periodic orbits $x_{s,u}(t)$ never cross the
matching point $x=0$ of the two parabolic pieces of $V(x)$, i.e. we
require that
\begin{equation}
x_s(t) < 0 < x_u(t)
\label{5.6a}
\end{equation}
for all times $t$. One finds that
this property is granted if and only if the conditions
\begin{equation}
  A^{2} < (\lambda_{s,u}^{2} + \Omega^{2})\, \xsu^{2}
  \label{5.7}
\end{equation}
are satisfied for both the `$s$'- and the `$u$'-indices, 
and that the periodic orbits then take the explicit form 
\begin{equation}
x_{s,u}(t) = \xsu - 
\frac{A\, [\lambda_{s,u}\, \sin(\Omega\, t)+\Omega\,\cos(\Omega\, t)]}
{\lambda_{s,u}^2+\Omega^2}\ .
\label{5.8}
\end{equation}
With the definitions (\ref{4.12},\ref{4.21},\ref{4.23}) it follows that 
\begin{eqnarray}
  \J_{s,u}(t,\tc) &=& \lambda_{s,u} \cdot (t-\tc) \label{5.9}\\
  \A_{s,u}(t)  &=& |\lambda_{s,u}|^{-1}\ .\label{5.10}
\end{eqnarray}

Our next goal is the determination of the master path
$x^{\ast}_{\opt}(t)$. 
To simplify the analytical calculations we restrict
ourselves to the case that the master path
$x^{\ast}_{\opt}(t)$ crosses the point $x=0$ exactly once, say at the time
$t=\tc$:
\begin{equation}
  x^{\ast}_{\opt}(t) = 0 \quad \Longleftrightarrow \quad t=\tc
  \label{5.11}
\end{equation}
The self-consistency of this assumption with the final
solution for $x^{\ast}_{\opt}(t)$ remains to be checked later.

 From (\ref{3.22},\ref{3.23}) we see that both $x^{\ast}_{\opt}(t)$ and
$p^{\ast}_{\opt}(t)$ are still continuous at $t=\tc$. For all other times
$t$ the relation (\ref{4.5}) and the hence following conclusions are 
not approximations but exact
identities since the force-field $F(x,t)$ in (\ref{5.4}) is by construction
piecewise linear. By introducing (\ref{5.10},\ref{5.11}) and (\ref{4.9}) into
(\ref{4.25},\ref{4.28}) we obtain by letting
$t_{0}\to\-\infty$ and $t_{f}\to\infty$ for the master 
path the following two relations 
(one with index `$s$' and one with `$u$'):
\begin{equation}
  x_{s,u}(\tc) = p^{\ast}_{\opt}(\tc)/\lambda_{s,u}\ .
  \label{5.12}
\end{equation}
These two equations for the two unknowns $\tc$ and $p^{\ast}_{\opt}(\tc)$
imply with (\ref{5.8}) the result
\begin{eqnarray}
  \tan(\Omega\, \tc) &=& \frac{1}{\Omega} 
                           \frac{\lambda_{s}\, \lambda_{u} - \Omega^{2}}
                               {\lambda_{s}+\lambda_{u}} \label{5.13} \\
   p^{\ast}_{\opt}(\tc) &=& \lambda_{u} \xu - 
                              \frac{A\, \lambda_{s}\, \lambda_{u}\cos(\Omega\, \tc)}
                                   {\Omega (\lambda_{s} + \lambda_{u})}  \label{5.14}
\end{eqnarray}
We observe that the solutions $\tc$ of (\ref{5.13}) are independent of
$A$. Furthermore, there are obviously two solutions $\tc$ within every
time-period $\T=2\pi/\Omega$. We anticipate, that only one of them corresponds
to a minimum of the action, and thus to the master path. Hence we fix $\tc$
uniquely (up to the usual degeneracy under $t\mapsto t+\T$) by (\ref{5.13}) 
in conjunction with 
\begin{equation}
  \frac{A\, \Omega\cos(\Omega\, \tc)}{\lambda_{s}+\lambda_{u}} < 0\ ,
  \label{5.15}
\end{equation}
and show later, that this condition singles out the right solution of (\ref{5.13}).
[The case $\lambda_{s}+\lambda_{u}=0$ has to be treated as limit
$\lambda_{s}+\lambda_{u}\to 0$.] Combining (\ref{5.13}-\ref{5.15}) it follows
that 
\begin{eqnarray}
  p^{\ast}_{\opt}(\tc) &=& \lambda_{u}\xu - 
                          |A|\, |\lambda_{s}|\,\lambda_{u}/\nu^{2}\, >\, 0 \ ,
                          \label{5.16}
\end{eqnarray}            
where we have introduced the definition              
\begin{eqnarray}          
  \nu^{2} &:=& [(\lambda_{s}^{2}+\Omega^{2})(\lambda_{u}^{2}+\Omega^{2})]^{1/2}
                        \ .
                        \label{5.17}
\end{eqnarray}
Note that $\lambda_{u}\xu$ in (\ref{5.16}) may be rewritten in various equivalent
forms according to
(\ref{5.6}) and that the last relation $p^{\ast}_{\opt}(\tc)>0$
in (\ref{5.16}) follows as a consequence of 
(\ref{5.3},\ref{5.6a},\ref{5.12}).

Given $\tc$ and $p^{\ast}_{\opt}(\tc)$, the entire time evolution of the
master path $x^{\ast}_{\opt}(t)$ can be readily inferred from
(\ref{4.9},\ref{4.10},\ref{4.21},\ref{4.23},\ref{4.25},\ref{4.28}) and
(\ref{5.9},\ref{5.10}) with the result
\begin{eqnarray}
  p^{\ast}_{\opt}(t) &=& p^{\ast}_{\opt}(\tc)\,
                         e^{-\lambda_{s,u}\cdot(t-\tc)}  
     \label{5.18}\\
  x^{\ast}_{\opt}(t) &=& x_{s,u}(t) - p^{\ast}_{\opt}(t)/\lambda_{{s,u}} 
     \label{5.19}
\end{eqnarray}
where `$s$' is associated with times $t\le \tc$ and `$u$' with $t\ge \tc$. All
the general qualitative features discussed in Sect.~IV.A.\ are nicely
illustrated by this explicit example (\ref{5.18},\ref{5.19}). 

Finally, we have to check the self-consistency of the solution
(\ref{5.18},\ref{5.19}) with our initial assumption (\ref{5.11}), 
i.e., we have to verify that $x^{\ast}_{\opt}(t)$ is strictly positive 
for $t>\tc$ and negative for $t<\tc$.
In general, in doing so,
a transcendental equation arises which has to be evaluated
numerically. Without going into the details of the proof we
further mention that one can show analytically that
%(\ref{5.7}) is a necessary but not 
%sufficient self-consistency criterion, while
$A^2 < \lambda_{s,u}^2\,\bar x_{s,u}^2$ 
is a sufficient but not necessary self-consistency criterion for (\ref{5.11}).
On the other hand, it is obvious that the
%(\ref{5.11}) obviously implies (\ref{5.6a}). is sufficient but not necessary. 
%Since (\ref{5.6a}) is equivalent to (\ref{5.7}), we see 
assumption $x_{s}(t)<0<x_{u}(t)$ in (\ref{5.6a}) is 
automatically covered by the stronger requirement (\ref{5.11}).
Thus, (\ref{5.6a},\ref{5.7}) is a necessary but not sufficient 
self-consistency criterion for (\ref{5.11}).

Introducing the above relations (\ref{5.18},\ref{5.19})
into (\ref{3.24},\ref{4.4}), we obtain for the action of the master path
\begin{equation}
  \phi_{\opt} = \Delta V \left[
                           1-\left|
                                \frac{A^{2}\, \lambda_{s}\, \lambda_{u}\,
                                      (|\lambda_{s}|+\lambda_{u})}
                                     {2\, \Delta V\, \nu^{4}}
                             \right|^{1/2}
                         \right]^{2}\ .\label{5.20}
\end{equation}
For $A\to 0$ or $\Omega\to\infty$ we thus recover the static (undriven)
potential barrier
$\Delta V$ from (\ref{5.5}). The leading order corrections for small $A$
decrease like $|A|$ \cite{sme99}. For any finite amplitude $A$ and driving
period $\T=2\pi/\Omega$ the ``effective potential barrier'' $\phi_{\opt}$ is
smaller than the static barrier $\Delta V$ and is monotonically decreasing
both with increasing $A$ and increasing $\T$. 
Invoking the necessary but not sufficient self-consistency criterion 
(\ref{5.7}) for (\ref{5.11}), one can explicitly confirm that
$\phi_{\opt}$ can never become zero (see (\ref{3.17},\ref{3.18},\ref{4.4}))
by demonstrating that the argument in the square-brackets in 
(\ref{5.20}) is always positive.
If we had chosen the solution of
(\ref{5.13}) with the opposite inequality than in (\ref{5.15}), then a plus
instead of the minus
sign in (\ref{5.18}) would have been the consequence. Thus (\ref{5.15}) is
indeed the pertinent condition for singling out the solution which
minimizes the action.

By using (\ref{5.9}) and (\ref{5.18}) in the definition (\ref{4.60'}) of
$\beta_{\opt}(t)$ we obtain 
\begin{equation}
  \beta_{\opt}(t) = p^{\ast}_{\opt}(\tc)\, e^{-\lambda_{u}\cdot(t-\tc)}  \ .
  \label{5.21}
\end{equation}
Turning to the prefactor $Q^{\ast}_{\opt}(t)$, we see from (\ref{4.41}) and
(\ref{5.10}) that
\begin{equation}
  Q^{\ast}_{\opt}(t) = \frac{1}{2\, |\lambda_{s}|}
  \label{5.22}
\end{equation}
for all times $t\le \tc$. Since $F''(x,t) = (\lambda_{u}-\lambda_{s})
\delta(x)$ according to (\ref{5.4}), we can infer from (\ref{3.27}) that the
prefactor $Q^{\ast}_{\opt}(t)$ is continuous at $t=\tc$ while its derivative
jumps from $\dot{Q}^{\ast}_{\opt}(t^{-}_{1}) = 0$ to the value
\begin{equation}
  \dot{Q}^{\ast}_{\opt}(t^{+}_{1}) = 
    \frac{|\lambda_{s}| + \lambda_{u}}{|\lambda_{s}|}\,
    \frac{\dot{x}^{\ast}_{\opt}(\tc)-p^{\ast}_{\opt}(\tc)}
         {\dot{x}^{\ast}_{\opt}(\tc)}\ ,
  \label{5.23}
\end{equation}
where $t^{+}_{1}$ indicates the limit $t\to \tc$ from above and $t^{-}_{1}$
from below. With these initial conditions, the solution of (\ref{4.39}) in the
domain $t>\tc$ is straightforward, yielding 
\begin{equation}
  \lim_{t\to\infty}Q^{\ast}_{\opt}(t)\, p^{\ast}_{\opt}(t)^{2} =
  \frac{p^{\ast}_{\opt}(\tc)^{2}\, \dot{Q}^{\ast}_{\opt}(t^{+}_{1})}
       {2\, \lambda_{u}}\ .
  \label{5.24}
\end{equation}
Using (\ref{5.14},\ref{5.18},\ref{5.19}) one can show that
$p^{\ast}_{\opt}(\tc) - \dot{x}^{\ast}_{\opt}(\tc)$ is identical to the
expression on the left hand side of (\ref{5.15}), so that (\ref{5.23}) and
thus (\ref{5.24}) are positive quantities. With the opposite inequality in
(\ref{5.15}) they would be negative, confirming once more that the latter 
case corresponds to a saddle point rather than a minimum of the action. 
Collecting everything, we are
finally in the position to evaluate eq. (\ref{4.52}) with the result
\begin{equation}
  \alpha_{\opt} = \left[
                    \frac{|A|\,(\Omega^{2}+\lambda_{s}\lambda_{u})+
                          \sqrt{\frac{2 \Delta V \nu^{4}}
                                     {|\lambda_{s}|^{-1}+{\lambda_{u}}^{-1}}}}
                         {16\, \pi^{3}\, |A|\, \phi_{\opt}}
                  \right]^{1/2}
  \label{5.25}
\end{equation}
Once again, the fact that the argument in the square root is positive can be
explicitly verified by exploiting the necessary but not sufficient self-consistency
criterion (\ref{5.7}) for (\ref{5.11}).

\subsection{Comparison of analytical and numerical results}

We have compared the above analytical predictions for the instantaneous
rate (\ref{4.51})
with very accurate numerical results in fig.4. 
To this end, we have computed the solution of the
Fokker-Planck-equation (\ref{2.17}) and then evaluated the rate according to
(\ref{2.20}), starting with an narrow Gaussian initial distribution $p(x,t_{0})$ 
about the potential well $\xs$ and then waiting until transients
have died out, i.e. until $\Gamma(t)$ has reached its $\T$-periodic asymptotic
behavior. 
%To solve the one-dimensional time-dependent Fokker-Planck-equation
%(\ref{2.17}) we have used the numerical routine ``{\tt D03PDF}'' of the NAG
%numerical library \cite{NAG}.  
In order to numerically evolve the one-dimensional time-dependent 
Fokker-Planck-equation (\ref{2.17}) one can employ standard parabolic
partial differential equation solving procedures in one spatial variable.
We have adopted a Chebyshev collocation method \cite{colloc}
%and the method of lines 
to reduce the problem to a coupled
system of ordinary differential equations, which
is then solved by standard numerical methods.
By changing the various parameters of the numerical procedure,
the typical relative errors of the numerical rates $\Gamma (t)$
in our figures are estimated to be at most of the order of $10^{-4}$ 
for rates down to about $10^{-100}$ and of the order of $10^{-3}$ 
for rates down to about $10^{-200}$.

The results in fig.4 confirm for a representative
set of parameter values that the agreement between
the analytical predictions and the practically exact numerical
results for the instantaneous rate $\Gamma(t)$
indeed improves with decreasing noise-strength
$D$. While the absolute values of $\Gamma(t)$ and the location of the
extrema strongly depend on $D$, the overall shape changes very little
and does {\em not} develop singularities as $D\to 0$.

The corresponding time averaged rates (\ref{4.60}) are depicted in
fig.5, exhibiting excellent agreement between theory and numerics even
for relatively large $D$.  The inset of fig.~5 confirms our 
analytical prediction
that the relative error in (\ref{4.60}) decreases asymptotically
like $D$, see (\ref{4.64}).

Finally, fig.6. illustrates the dependence of the time averaged rate
$\bar\Gamma$ upon the amplitude $A$ of the periodic driving force.  
As expected, our theoretical prediction compares very well with the
(numerically) exact rate, except for very small driving amplitudes
$A$. The latter discrepancy is in accordance with our
discussion in Sect.~IV.A and Sect.~IV.E.  

We have furthermore included in fig.6 a comparison with the analytical
approximation for the time averaged rate $\bar\Gamma$
from Ref. \cite{sme99}. By way of a matching procedure, 
involving the barrier region only, it is predicted \cite{sme99} that
\begin{equation}
  \bar\Gamma = \Gamma_{0} \int_{0}^{2\pi}\frac{d\phi}{2\pi}\, e^{-s(\phi)/D}
  \ ,\label{5.26}
\end{equation}
where $\Gamma_{0}$ is the well-known Kramers-Smoluchowsky-rate in the absence
of the periodic driving force \cite{han90}
\begin{equation}
  \Gamma_{0} = \frac{|V''(\xs)\, V''(\xu)|}{2\,\pi}^{1/2}\, e^{-\Delta V/D}\ .
  \label{5.27}
\end{equation}
The leading order effect of an additive sinusoidal driving (\ref{2.5}), such
that the associated periodic modulations of the potential barrier are small in
comparison with the unperturbed barrier $\Delta V$, but not necessarily in
comparison with the noise-strength $D$, are captured by the function $s(\phi)$
in (\ref{5.27}). It can be written for a general metastable potential $V(x)$
in (\ref{2.5}) under the form \cite{sme99} 
\begin{eqnarray}
  s(\phi) &=& A\, (S\, \sin\phi - C\, \cos\phi) \label{5.28} \\
  S &=& S(x_{1}) := \int_{\xs}^{\xu}dx \sin\left(\Omega \int_{x_{1}}^{x}\frac{dy}{V'(y)}\right)\nonumber\\
  C &=& C(x_{1}) := \int_{\xs}^{\xu}dx \cos\left(\Omega \int_{x_{1}}^{x}\frac{dy}{V'(y)}\right)\label{5.29}
\end{eqnarray}
with an arbitrary reference point $x_{1}\in(\xs,\xu)$.
Fig.6 confirms that this approximation from \cite{sme99}
is indeed complementary to ours in that it 
%admits, for a fixed (small) noise 
%strength $D$, arbitrarily small driving amplitudes $A$
is very accurate for small driving amplitudes
$A$ but develops considerable deviations with increasing $A$.  
Those approximations have been omitted in figs.4 and 5 since they are not
valid in this parameter regime and indeed are way off.

\subsection{Cubic potential}

As a second example we consider a force field~(\ref{2.5}) with a
cubic metastable potential
\begin{equation}
  V(x) = - \frac{a}{3}\,x^{3} + \frac{b}{2}\, x^{2} \ , \ \ a,b>0 \ .
%+ c\, x.
  \label{5.30}
\end{equation}
The stable and unstable fixed points $\xsu$ of this potential are given by
\begin{equation}
  \xs = 0 \ \ , \ \ \  \xu = \frac{b}{a},  \label{5.31}
\end{equation}
with curvatures at those fixed points 
\begin{equation}
  V''(\xs) = b\ \ , \ \ \ V''(\xu) = -b,  \label{5.32}
\end{equation}
and a static potential barrier height 
\begin{equation}
  \Delta V := V(\xu) - V(\xs) = \frac{b^{3}}{6a^{2}}.  \label{5.33}
\end{equation}
The time-dependent force field~(\ref{2.5}) takes the following form:
\begin{equation}
  F(x,t) = a\,x^{2} - b\,x 
%+ c 
  + A\,\sin(\Omega\,t)\ .  
  \label{5.34}
\end{equation}

Since already the analytical evaluation of such a force field's
periodic orbits is impossible, one has to recourse to numerical methods for the
calculation of the quantities $\phi_{\opt}$, $\alpha_{\opt}$, and
$\kappa_{\opt}(t,D)$ appearing in the rate expressions (\ref{4.51},\ref{4.60}). 
A convenient numerical strategy for doing so
has been discussed in detail already in Sect.~IV.H.
%We therefore follow our numerical recipe from
%Sect.~IV.G, testing as in the prevision section 
The so obtained predictions for the time averaged rate~(\ref{4.60}) 
are compared in fig.7 against precise numerical results 
%from an evolution of the Fokker-Planck-equation~(\ref{2.17}). The outcome of these calculations 
for a representative set of parameter values.
% is shown in fig.6,
Note that these parameter values are quantitatively
very similar to those in fig.6, hence also the rates as a function of
the noise strength $D$ are very similar.
The agreement between the theoretical prediction and the practically
exact numerical results is again excellent even for relatively large noise 
strengths $D$.
However, in contrast to the piecewise
parabolic case, 
the numerically accessible $D$-values are still not
small enough in order to check the validity of our
prediction (\ref{4.64}) for the behavior of the relative 
error in the analytical approximation (\ref{4.60}).

\section{Conclusions}
In our present work we have scrutinized by means of path
integral methods the thermally activated escape of an
overdamped Brownian particle over a periodically oscillating 
potential barrier in the most challenging regime of weak 
thermal noise in combination with moderately strong and 
moderately fast driving.

A first major result of our novel path-integral approach
is the expression (\ref{4.1}) for the instantaneous escape 
rate, which displays the suggestive general structure of
``probability at the separatrix times velocity''.
The summation appearing in this expression reflects the fact
that several local minima of the relevant action in the 
path-integral formulation of the escape problem notably 
contribute to the rate. In contrast to the undriven escape problem,
giving rise to a (quasi-) Goldstone mode due to the (quasi-) 
time-translation symmetry, in our present case the paths
corresponding to the local minima of the action are well separated
and therefore admit a standard saddle point approximation
of the path-integral for small noise strengths $D$.
Pictorially speaking, by switching on the periodic driving,
the continuous time-translation symmetry of the escape problem
is reduced to a time-discrete one.

Our present explorations indicate that from the
practical (technical) viewpoint, a path-integral approach which
keeps an entire sum of possibly relevant contributions to the
rate may be easier to handle than WKB-type or quasipotential-type
methods \cite{gra85,mai97}, which operate with the concept of a single
exponentially dominating weak-noise contribution and a single
pre-exponential factor, both of them typically of a non-analytic
nature.

The central result of our present paper represents the formula
(\ref{4.51}) for the instantaneous rate, supplemented by the
result (\ref{4.64}) for the ``accuracy exponent'' $\gamma$.
The above discussed summation over the relevant local minima of the
action resurfaces in all the equivalent
alternative expressions (\ref{4.53},\ref{4.53'},\ref{4.63}) for
$\kappa_{{\rm opt}}(t,D)$ but drops out (can be performed)
in the time averaged rate (\ref{4.60}) due to the miraculous identity 
(\ref{4.57}).

The rate expressions (\ref{4.51},\ref{4.60}) share the general 
Arrhenius-type structure of the exponentially
leading weak-noise contribution with the typical form of an 
equilibrium (undriven) rate (\ref{5.27}).
However, both, the Arrhenius factor and the pre-exponential contribution
to the rates (\ref{4.51},\ref{4.60}) now depend in a very complicated way
on global features of the periodically oscillating potential
(in contrast to the purely local properties $\Delta V = V(\xu)-V(\xs)$ and 
$V''(\xsu)$ governing (\ref{5.27})).
Moreover, a non-trivial $\sqrt{D}$-dependence of the pre-exponential
factor on the noise strength $D$ arises.

For the time averaged rate (\ref{4.60}) we have
shown in Appendix D that for asymptotically
weak noise $D$ an invariance property
holds under the supersymmetry transformation
(\ref{susy1}) without any further restrictions on 
the force field $F(x,t)$.
Such an invariance property
can be established rigorously on very general grounds 
\cite{jun91} for force fields of the
form $F(x,t)=-V'(x)+y(t)$ and arbitrary noise-strengths $D$.

The time averaged rate (\ref{4.60}) displays a remarkable 
structural similarity with the rate-expressions obtained in \cite{rei94}
for one-dimensional discrete-time systems in the presence of 
weak Gaussian white noise.
While a general qualitative connection between these two different
types of escape problems via some kind of stroboscopic mapping
is quite suggestive, the quantitative details are not so simple.
Especially, the Gaussianity of the resulting noise after the stroboscopic 
mapping is crucial \cite{rei96} but is far from obvious \cite{rei95}
for the rare but strong fluctuations (large deviations) which 
govern the escape events.

The conditions for the validity of our rate-formulas are (\ref{4.68}) and
that for a fixed (small) noise strength $D$, extremely weak, fast, and slow
periodic driving forces should be excluded.
Especially, the weak noise limit $D\to 0$ displays a rather intriguing 
non-interchangeability with the long-time limit $t\to\infty$ (see Sect.~IV.F), 
and with the limits of asymptotically weak, fast, or slow driving.

In general, an action-integral remains to be minimized numerically and 
an ordinary linear differential equation of second order for the prefactor
to be solved (Sect.~IV.H) before actual numbers can be extracted from our 
rate-formulas.
However, for the special case of a sinusoidally driven, piecewise 
parabolic metastable potential this entire program can be executed 
in closed analytical from. This example retains all the typical
features of more general setups and exhibits excellent agreement with
high-precision numerical results (Sect.~V).

Conceptionally, our new path-integral approach should be of 
considerable interest for many related problems.  
Generalizations for higher dimensional systems and for
non-periodic driving forces are currently under investigation \cite{leh99}.
Also the proper handling of the tantalizing
weak, fast, and slow driving limits within a consistent path-integral 
formalism remains as an open problem for future research.

Finally, the complicated dependence of the rate on the global
details of the oscillating potential poses an challenging
inverse problem, namely to reconstruct the underlying
force field from a given (e.g. measured) behavior of the
escape rates.

\subsection*{Acknowledgement}
This work has been supported by the DFG-Sachbeihilfe HA1517/13-2 (P.H., P.R.), the
Sonderforschungsbereich 486 (P.H., J.L.), the Studienstiftung des deutschen
Volkes (J.L.), and the
Graduiertenkolleg GRK283 (P.H., P.R.).

\section*{Appendix A}
To prove the equivalence of (\ref{3.11}) and (\ref{3.12}-\ref{3.14}) one first
needs the Hessian of the discrete-time action $S_{N}(x_{0},\dots,x_{N})$ in
(\ref{3.7}), which is given by the $(N-1)\times(N-1)$-matrix
%\begin{equation}
%  \frac{\partial^2 S({\bf x}^\ast)}{\partial x_n^\ast\partial x_m^\ast} = 
%  \frac{1}{2\Delta t}\,\left(a_{n}\,\delta_{n,m}  - b_{n}\,\delta_{n+1,m} -
%       b_{m}\,\delta_{n-1,m}\right)\ , \label{a1}
%\end{equation}
\begin{eqnarray}
 & & \left(2 \Delta t\,\frac{\partial^2 S({\bf x}^\ast)}{\partial x_n^\ast\partial
 x_m^\ast}\right)_{n,m=1,2,...,N-1} = \nonumber \\
 & & = \left(
    \begin{array}{ccccc}
      a_{1}  & - b_{1} &        &           & \\
      -b_{1} & a_{2}   & \smash{\ddots} &           &  \\
             & \ddots  & \ddots &  a_{N-2}  & -b_{N-2} \\
             &         &        &  -b_{N-2} & a_{N-1}
    \end{array}
  \right)\ , \label{a1}
\end{eqnarray}
where
\begin{eqnarray}
 a_{n} &:= 2 &\, + \big[ 
                     \,2 \,F'(x^{\ast}_{n},t_{n}) -
                     \left(x^{\ast}_{n+1}-x^{\ast}_{n}\right) F''(x^{\ast}_{n},t_{n})\,
                  \big] \Delta t \nonumber\\
            & &\, + \big[ 
                   \left(F'(x^{\ast}_{n},t_{n})\right)^{2} +
                   F(x^{\ast}_{n},t_{n})\, F''(x^{\ast}_{n},t_{n})\,
                 \big] \Delta t^{2} \label{a2} \\
 b_{n} &:= 1 &\, +\, F'(x^{\ast}_{n},t_{n})\,\Delta t\ . \label{a3}
\end{eqnarray}
For the prefactor $Z_{N}({\bf x}^{\ast})$ in (\ref{3.11}) the determinant of
the matrix on the right-hand side of (\ref{a1}) has to be evaluated. This can
be done by a standard procedure (cf. \cite{sch81}).
The result is a linear two-step
recursion relation for the principal minor $\tilde{Q}^{\ast}_{n}$, 
consisting of the first $n$ columns and rows of (\ref{a1}), of the form
\begin{equation}
  \tilde{Q}^{\ast}_{n+1} =   a_{n}\, \tilde{Q}^{\ast}_{n} 
                           - b_{n-1}^{2}\, \tilde{Q}^{\ast}_{n-1}
%  \quad
%  (1<n<N)
  \label{a4}
\end{equation}
for $2\leq n\leq N$, with initial conditions
\begin{equation}
  \tilde{Q}^{\ast}_{1} = 1\ ,\quad \tilde{Q}^{\ast}_{2} = a_{1}\ .
  \label{a5}
\end{equation}
Comparing (\ref{3.11},\ref{3.14}) with (\ref{a1}), we observe that
$Q^{\ast}_{N} = \Delta t \, \tilde{Q}^{\ast}_{N}$, and due to the linearity of
(\ref{a4}) we can conclude that $Q^{\ast}_{n}:=\Delta t\,
\tilde{Q}^{\ast}_{n}$ also obeys these equations.  Therefore,
using the definitions~(\ref{a2},\ref{a3}}), it is readily shown that $Q^{\ast}_{n}$ satisfies
the required recursion relations~(\ref{3.12}) together with the initial
conditions~(\ref{3.13}).

As a by-product, needed in Appendix B, we notice that by defining  for $1\le n
< N$
\begin{equation}
  \mu_{n}:=\frac{\tilde{Q}^{\ast}_{n+1}}{\tilde{Q}^{\ast}_{n}}
  =\frac{Q^{\ast}_{n+1}}{Q^{\ast}_{n}} \ ,
  \label{a6}
\end{equation}
the linear two-step recursion relation (\ref{a4}) can be rewritten as an 
equivalent non-linear one-step recursion
\begin{equation}
  \mu_{n+1} = a_{n+1} - \frac{b_{k}^{2}}{\mu_{n}}\ ,\quad \mu_{1}=a_{1}\ .
  \label{a7}
\end{equation}
The actual quantity of interest
$\tilde{Q}^{\ast}_{N}$ then follows as
\begin{equation}
  \tilde{Q}^{\ast}_{N} = \prod_{n=1}^{N-1} \mu_{n}\ .
  \label{a8}
\end{equation}

\section*{Appendix B}
In the following we derive Eq.~(\ref{3.31}) by showing that 
\begin{equation}
  \frac{\partial^{2} \phi_{k}(x_{f},t_{f})}{\partial x_{f}^{2}} = 
  \frac{\dot{Q}^{\ast}_{k}(t_{f})}{2\, Q^{\ast}_{k}(t_{f})} - F'(x_{f},t_{f})\ ,
  \label{b1}
\end{equation}
which gives together with (\ref{3.20},\ref{3.29}) for $t=t_{f}$ the desired result.

We work with the time-discrete version of the quantities in (\ref{b1}) and
consider in a first step the dependency of a minimizing path 
${\bf x}^{\ast}= {\bf x}^{\ast}(x_{f})$
on the end-point $x_{f}$ for fixed $t_0$, $t_{f}$, and $x_{0}$. 
In order to complicate the notation as little as
possible, we have left out the index $k$ labeling the different paths 
${\bf  x}^{\ast}_{k}$. 

Since the initial point $x_0$ is kept fixed, we have that $dx_0^\ast/dx_f = 0$.
Further, we know
from (\ref{3.8}) that for all $n=1,\dots,N-1$
\begin{equation}
  \frac{\partial S_{N}}{\partial x_{n}}\bigg|_{{\bf x}^{\ast}(x_{f})} \!\!= 0\ ,
  \label{b2}
\end{equation}
for any $x_f$-value,
which implies after taking the derivative with respect to~$x_{f}$ that
\begin{equation}
  \sum_{m=1}^{N-1} 
     \frac{dx^{\ast}_{m}}{dx_{f}}\,
     \frac{\partial^{2} S_{N}}{\partial x_{m} \partial x_{n}}
     \bigg|_{{\bf x}^{\ast}(x_{f})}\!\!
  = 0\ .
  \label{b3}
\end{equation}
Using (\ref{a1}), we thus get:
\begin{equation}
    a_{n}  \,\frac{dx^{\ast}_{n}}{dx_{f}} 
  - b_{n-1}\,\frac{dx^{\ast}_{n-1}}{dx_{f}}
  - b_{n}  \,\frac{dx^{\ast}_{n+1}}{dx_{f}} = 0
  \label{b4}
\end{equation}
Introducing the new quantities $\eta_{n}$ by
\begin{equation}
  \eta_{n} := b_{n}\,\frac{dx^{\ast}_{n+1}}{dx^{\ast}_{n}}
  \label{b5}
\end{equation}
one obtains from (\ref{b4}) the one-step recursion relation
\begin{equation}
  \eta_{n+1} = a_{n+1} - \frac{b_{n}^{2}}{\eta_{n}}
  \label{b6}
\end{equation}
The corresponding initial condition follows from (\ref{b4}) for $n=1$ by taking
into account the above mentioned fact that $dx^{\ast}_{0}/dx_{f} = 0$:
\begin{equation}
  \eta_{1}=a_{1}\ .
  \label{b7}
\end{equation}
Comparing (\ref{a7}) with (\ref{b6},\ref{b7}) yields $\eta_{n}=\mu_{n}$ for
$n=1,\dots,N-1$ and therefore, using the definitions 
(\ref{a3},\ref{a6},\ref{b5}), for $n=N-1$
\begin{equation}
  \frac{Q^{\ast}_{N}}{Q^{\ast}_{N-1}} = 
  \left[1+F'(x^{\ast}_{N-1},t_{N-1})\right] \frac{dx^{\ast}_{N}}{dx^{\ast}_{N-1}}\ .
  \label{b8}
\end{equation}

In a next step an explicit expression for $dx^{\ast}_{N}/dx^{\ast}_{N-1}$ in
terms of well-known quantities has to be found. This can be achieved by taking the
second derivative of the discrete-time action $S_{N}({\bf x}^{\ast}(x_{f}))$
of the same minimizing path ${\bf x}^{\ast}$ as above with respect to the
end-point $x_{f}$. With (\ref{b2},\ref{b3}), and the boundary
conditions (\ref{3.7a}) we find
\begin{equation}
  \frac{d^{2}S_{N}({\bf x}^{\ast}(x_{f}))}{dx_{f}^{2}} =
  \sum_{n=1}^{N} \frac{d x^{\ast}_{n}}{dx^{\ast}_{N}}
  \frac{\partial^{2} S_{N}}{\partial x_{n} \partial x_{N}}\bigg|_{{\bf
  x}^{\ast}(x_{f})}\ ,
  \label{b9}
\end{equation}
and using (\ref{3.7}) we can conclude that
\begin{equation}
  \frac{dx^{\ast}_{N}}{dx^{\ast}_{N-1}} = 
  \frac{1 + \Delta t\, F'(x^{\ast}_{N-1},t_{N-1})}
       {1-2\,\Delta t\,\frac{d^{2}}{dx_{f}^{2}}\, S_{N}({\bf x}^{\ast}(x_{f}))}
  \ .
  \label{b10}
\end{equation}
Together with (\ref{b8}) we thus arrive at 
\begin{equation}
  \frac{Q^{\ast}_{N}}{Q^{\ast}_{N-1}} = 
  \frac{\left[1+\Delta t\,F'(x^{\ast}_{N-1},t_{N-1})\right]^{2}}
       {1-2\,\Delta t\,\frac{d^{2}}{dx_{f}^{2}}\, S_{N}({\bf x}^{\ast}(x_{f}))}
  \ ,
  \label{b11}
\end{equation}
which with (\ref{3.24}) yields in the continuous-time limit the searched for
relation (\ref{b1}).

\section*{Appendix C}
In this appendix we derive (\ref{4.45'}) for $t\in[t_u,t_f]$, where the order of
magnitude refers to the asymptotics with respect to $k$.  

As discussed below (\ref{4.32}), the differences
\begin{eqnarray}
\delta x^\ast_k(t) & := & x_k^\ast(t) - x_\opt^\ast(t+k\T)\label{c1}\\
\delta p^\ast_k(t) & := & p_k^\ast(t) - p_\opt^\ast(t+k\T)\label{c2}
\end{eqnarray}
rapidly decrease with increasing index $k$ uniformly on the entire time-interval
$[t_0,t_f]$.
Our first conclusion is
that the time-point $t_u$ at which $x^\ast_u(t)$ enters the neighborhood of
$x_u(t)$ depends itself on the index $k$, basically it decreases like $-k\,\T$,
see fig.3. On the other hand, the distance $\dx_k(t_u)$ 
at which the path enters this neighborhood is, by definition, $k$-independent.
The corresponding momentum $p^\ast_k(t_u)$ is not strictly $k$-independent,
but approaches an asymptotic $k$-independence for large $k$.
Similar conclusions  apply for the time $t_s$ at which $x^\ast_u(t)$ leaves the
neighborhood of the stable periodic orbit. 

%To gain further quantitative insight,
%we note that (\ref{4.10}) implies
%\begin{equation}
%\delta p^\ast_k(t) = \delta p^\ast_k(t_{s,u})\, e^{-\J_{s,u}(t,t_{s,u})}
%\label{ca}
%\end{equation}
%whence the following $t$-independences can be inferred
%\begin{equation}
%\frac{\delta p^\ast_k(t)}{p^\ast_k(t)} = \mbox{const.}\ \ ,\ \ \
%\frac{\delta p^\ast_k(t)}{p^\ast_\opt(t+k\T)} = \mbox{const.}\ .
%\label{cb}
%\end{equation}
 
Next we can conclude from (\ref{4.20},\ref{4.25}) that
\begin{equation}
\delta x^\ast_k(t) = \delta p^\ast_k(t)\, \A_s(t) - 
p^\ast_k(t)\, e^{2\,\J_s(t,t_0)}\, \A_s(t_0)\ .
\label{c3}
\end{equation}
Within the approximation (\ref{4.32.3}) it follows that 
$\delta p^\ast_k(t_s) = \delta x^\ast_k(t_{s})/\A_s(t_s)$.
With these initial conditions at $t=t_s$, the small perturbations 
$\delta x^\ast_k(t)$ and $\delta p^\ast_k(t)$ are then propagated
according to (\ref{3.22},\ref{3.23}) until $t=t_u$. In linear order 
of these small perturbations it follows that 
$\delta p^\ast_k(t_u) / \delta x^\ast_k(t_{u})$ is an
asymptotically $k$-independent constant, which, however, depends on all
the details of the force field $F(x,t)$ along the crossover segment
of the master path $x^\ast_\opt (t)$. 

The counterpart of (\ref{c3}) in the neighborhood
of $x_u(t)$ follows along the same line of reasoning, reading
\begin{equation}
\delta x^\ast_k(t) = - \delta p^\ast_k(t)\, \A_u(t) +
p^\ast_k(t)\, e^{-2\,\J_s(t_f,t)}\, \A_u(t_f)\ .
\label{c3'}
\end{equation}
Replacing on the right hand side $e^{-2\,\J_s(t_f,t)}$ by
$p^\ast_\opt(t_f+k\T)^2/p^\ast_\opt(t+k\T)^2$ 
according to (\ref{4.10},\ref{4.16}), choosing $t=t_u$,
and making use of (\ref{4.32'},\ref{c2}) we can infer that
\begin{equation}
\delta x^\ast_k (t_u) + \delta p^\ast_k (t_u)\, \A_u(t_u) = 
\frac{p^\ast_k (t_u)\, \A_u(t_f)}{[p^\ast_k (t_u) -\delta p^\ast_k (t_u)]^2}
\, p^\ast_\opt(t_k)^2 \ .
\label{c4}
\end{equation}
As we have just pointed out, the quantity 
$\delta x^\ast_k(t_u)$ is proportional to 
$\delta p^\ast_k(t_u)$ with an asymptotically $k$-independent
proportionality constant that depends on
the details of the force field $F(x,t)$ along the crossover segment
of the master path $x^\ast_\opt (t)$. 
In the generic case, this proportionality constant is thus not
expected to coincide with $-\A_u(t_u)$ since the latter
depends on the behavior of
$F(x,t)$ along the unstable periodic orbit $x_u(t)$ only.
Consequently,
both $\delta x_k^\ast (t_u)$ and $\delta p_k^\ast (t_u)$ on 
the left hand side of (\ref{c4})
are, with respect to their $k$-dependence, of the same order 
of magnitude as the right hand side.
Since $p^\ast_k(t_u)$ is asymptotically $k$-independent and
$\delta p^\ast_k(t_u)$ tends to zero, we can infer from (\ref{c4})
that
\begin{equation}
\delta x^\ast_k (t_u)  = \Ord (p^\ast_\opt(t_k)^2)\  ,\ \
\delta p^\ast_k (t_u)  = \Ord (p^\ast_\opt(t_k)^2)\ .
\label{c5}
\end{equation}

With the initial conditions (\ref{4.42'},\ref{4.42''}) it follows from (\ref{3.27})
that the relative difference between $Q^\ast_k(t_u)$ and $Q^\ast_\opt(t_u+k\T)$ 
scales as a function of $k$ like 
$\delta x^\ast_k (t_u)$ and $\delta p^\ast_k (t_u)$. With (\ref{c5}) this implies that
\begin{equation}
Q^\ast_k(t_u) = Q^\ast_\opt(t_u+k\,\T)\, [1+\Ord (p^\ast_\opt(t_k)^2)]\ .
\label{c6}
\end{equation}
A similar relation follows for $\dot Q^\ast_k(t_u)$ and thus for $g^\ast_k(t_u)$
(cf. (\ref{3.29})), namely
\begin{equation}
g^\ast_k(t_u) = g^\ast_\opt(t_u+k\,\T)\, [1+\Ord (p^\ast_\opt(t_k)^2)]\ .
\label{c7}
\end{equation}
Finally we conclude from (\ref{4.45.1}) that
\begin{eqnarray}
\frac{Q^\ast_k(t)}{Q^\ast_\opt(t+k\T)} & = & 
\frac{Q^\ast_k(t_u)}{Q^\ast_\opt(t_u+k\T)}\nonumber\\
& \times & 
\frac{1-g^\ast_k(t_u)\, \A_u(t_u,t)}{1-g^\ast_\opt(t_u+k\T)\, \A_u(t_u,t)}
\label{c8}
\end{eqnarray}
Like for $\dx_k(t_u)$ and $p^\ast_k(t_u)$ (see below (\ref{c2})) 
one can convince oneself 
that also $g^\ast_k(t_u)$ is asymptotically $k$-independent.
With (\ref{c6},\ref{c7}) the result (\ref{4.45'}) 
then follows from ({\ref{c8}).

\section*{Appendix D}

The purpose of this appendix is to verify that our expression (\ref{4.60})
for the time-averaged rate is invariant with respect to the supersymmetry
transformation (\ref{susy1}). To this end, we first note that the path
defined via
\begin{eqnarray}
  \tilde{x}^{\ast}_{\opt}(t) & := & - x^{\ast}_{\opt}(-t)\ , 
  \label{d1}\\
  \tilde{p}^{\ast}_{\opt}(t) & := & p^{\ast}_{\opt}(-t)
  \label{d2}
\end{eqnarray}
satisfies the Hamilton equations (\ref{3.22},\ref{3.23}) for the
supersymmetric partner field $\tilde F(x,t)$ from eq.~(\ref{susy1}). Since
the periodic orbits of this new force field are given by $\tilde
x_s(t)= -x_{u}(-t)$ and $\tilde x_u(t) = -x_s(-t)$ (see Sect.~I.C)
one can readily see that $\tilde{x}^{\ast}_{\opt}(t)$ from (\ref{d1}) also
obeys the boundary conditions (\ref{4.3}) in the relevant limit
$t_{f}-t_{0}\to \infty$.  Hence we have found (up to the usual degeneracy with
respect to time shifts by arbitrary multiples of $\T$) the unique solution of
the supersymmetric partner variational problem~(\ref{4.4}). Inserting
$\tilde{p}^{\ast}_{\opt}(t)$ from (\ref{d2}) into the definitions (\ref{3.24})
and (\ref{4.4}) then leads to the following result:
\begin{equation}
  \tilde{\phi}_{\opt} = \phi_{\opt}\ .
  \label{d3}
\end{equation}

Somewhat more elaborate considerations are necessary in order to establish a
corresponding identity for the prefactor $\alpha_{\opt}$ in~(\ref{4.60}). To
this end we first consider two arbitrary but linear independent solutions
$Q_{i}(t)\ (i=1,2)$ of the pre\-factor equation (\ref{3.27}) for
$Q^{\ast}_{\opt}(t)$. One can then easily verify that the prefactor
$Q^{\ast}_{\opt}(t)$, which moreover has to fulfill the initial conditions
(\ref{3.28}) in the limit $t_{0}\to -\infty$, is given by
\begin{equation}
  Q^{\ast}_{\opt}(t) = 
  \lim_{t_{0}\to-\infty} 
    \frac{Q_{1}(t)\,Q_{2}(t_{0}) - Q_{1}(t_{0})\,Q_{2}(t)}
         {W(t_{0})}\ ,
  \label{d4}
\end{equation}
with the Wronskian 
\begin{equation}
  W(t) := \dot{Q}_{1}(t)\,Q_{2}(t) - Q_{1}(t)\,\dot{Q}_{2}(t)
  \ .
  \label{d5}
\end{equation}
Due to (\ref{3.27}) one can infer that
\begin{equation}
  \dot{W}(t) = 2 \, W(t)\, F'(x^{\ast}_{\opt}(t),t)\ .
  \label{d6}
\end{equation}
With help of the Hamilton equation~(\ref{3.22}) it follows that
\begin{equation}
  p^{\ast}_{\opt}(t)^{2} \, W(t) = {\rm const.}
  \label{d7}
\end{equation}

Turning now to the supersymmetric partner problem, it is readily seen that one
obtains via $\tilde{Q}_{i}(t):=Q_{i}(-t)$ two linear independent
solutions of the prefactor equation (\ref{3.27}) for the supersymmetric
partner field (\ref{susy1}) and the path given by (\ref{d1}).  Thus we can use
(\ref{d2}) and (\ref{d4},\ref{d5}) (with tildes) to establish the
identity
\begin{eqnarray}
  & &\lim_{t\to\infty} \tilde{p}^{\ast}_{\opt}(t)^{2} \,
  \tilde{Q}^{\ast}_{\opt}(t)
  \nonumber
  = 
  \lim_{t_{0}\to-\infty \atop t\to\infty} 
  \frac{p^{\ast}_{\opt}(-t)^{2}}
       {-W(-t_{0})} \\
  & & \times
  \left[
    Q_{1}(-t)\, Q_{2}(-t_{0}) - Q_{1}(-t_{0})\, Q_{2}(-t)
  \right]\ .
  \label{d8}
\end{eqnarray}
According to (\ref{d7}) we can now rewrite $p^{\ast}_{\opt}(-t)^{2}/W(-t_{0})$
as $p^{\ast}_{\opt}(-t_{0})^{2}/W(-t)$. 
Replacing $t\rightarrow -t_{0}$ and vice versa one can then conclude 
with help of (\ref{d4}) that
\begin{equation}
  \lim_{t\to\infty} \tilde{p}^{\ast}_{\opt}(t)^{2} \,
  \tilde{Q}^{\ast}_{\opt}(t) =
  \lim_{t\to\infty} p^{\ast}_{\opt}(t)^{2} \,
  Q^{\ast}_{\opt}(t)\ .
  \label{d9}
\end{equation}
Hence we finally obtain
\begin{equation}
  \label{d10}
  \alpha_{\opt} = \tilde{\alpha}_{\opt}\ .
\end{equation}
This, in combination with (\ref{d3}), proves our proposition that the time-averaged rate
(\ref{4.60}) is invariant with respect to the supersymmetric transformation
(\ref{susy1}).

\begin{figure}[htbp]
  \begin{center}
    \vspace*{-4.0cm} \epsfig{file=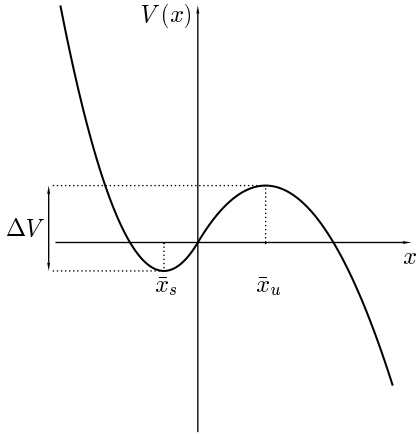,width=1.1\linewidth} \vspace*{-4.0cm}
  \end{center}              
  \caption{
    Sketch of a typical metastable potential $V(x)$ in eq.(\ref{2.5}).  
    Plotted is the piecewise parabolic example (\ref{5.1}) with 
    parameters $\Delta V = 0.9$, $\lambda_s = -0.6$, and $\lambda_u = 0.3$
    in arbitrary, dimensionless units.
    }
\end{figure}

\begin{figure}[htbp]
  \begin{center}
    \vspace*{-4.0cm} \epsfig{file=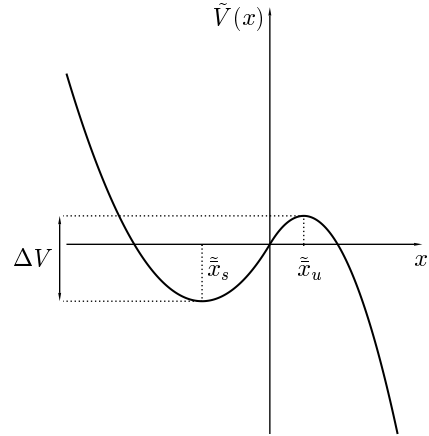,width=1.1\linewidth} \vspace*{-4.0cm}
  \end{center}              
  \caption{
    The supersymmetric partner potential $\tilde V(x) := - V(-x)$
    of the potential $V(x)$ from fig.1.
    }
\end{figure}

\begin{figure}[htbp]
  \begin{center}
    \vspace*{-1.0cm} \epsfig{file=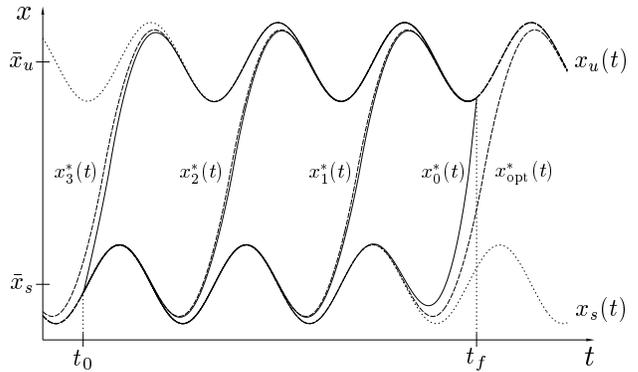,width=1.0\linewidth} \vspace*{-0.8cm}
  \end{center}              
  \caption{
    Solid: The paths $x^\ast_k(t)$, $k=0,\dots,K(t_f,t_0)=3$ which minimize the
    action (\ref{3.17},\ref{3.18}) with boundary conditions (\ref{4.3}).
    Dashed: The associated ``master paths'' $x^\ast_\opt(t+k\,\T)$, implicitly defined
    via (\ref{4.4}).
    Dotted: Stable and unstable periodic orbits $x_s(t)$ and $x_u(t)$ from (\ref{2.14a}).
    In this plot, $t_f-t_0$ has been chosen 
    rather small. As $t_f-t_0$ increases, more and more intermediate 
    paths $x^\ast_k(t)$ appear which better and better agree with their associated 
    master paths $x^\ast_{\opt}(t+k\T)$.
    The depicted curves have been obtained for the additively driven piecewise
    parabolic potential (\ref{2.5},\ref{5.1}) with parameters $A=0.5$,
    $\Omega=1$, $\lambda_s=-1$, $\lambda_u=1$, $\Delta V=1$,
    $t_0=-12$, $t_f=7.5$ (dimensionless units).
    }
\end{figure}

\begin{figure}[htbp]
  \begin{center}
    \vspace*{-4.8cm} \epsfig{file=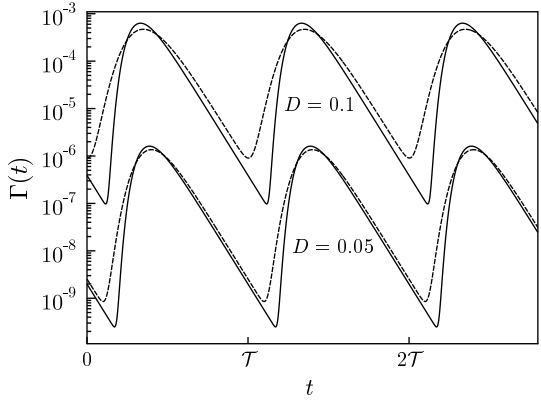,width=1.1\linewidth} \vspace*{-4.3cm}
  \end{center}              
  \caption{Instantaneous rate $\Gamma(t)$ versus time $t$
    for the piecewise linear force field
    (\ref{5.4}) with parameters  $x_{s}=\lambda_{s}=-1$,
    $x_{u}=\lambda_{u}=1$, $\Omega=1$, and $A=0.5$,
    corresponding to a static ($A=0$) potential barrier $\Delta V = 1$ 
    in (\ref{5.5},\ref{5.6}). 
    Solid line: Analytical prediction 
    (\ref{4.51},\ref{4.63},\ref{5.16},\ref{5.17},\ref{5.20},\ref{5.21},\ref{5.25}) 
    by neglecting the $\Ord (D^{\gamma})$-term in (\ref{4.51}). 
    Dashed line: High-precision 
    numerical results, obtained as described in Sect.~V.B.
   }
\end{figure}

\begin{figure}[htbp]
  \begin{center}
    \vspace*{-5.0cm} \epsfig{file=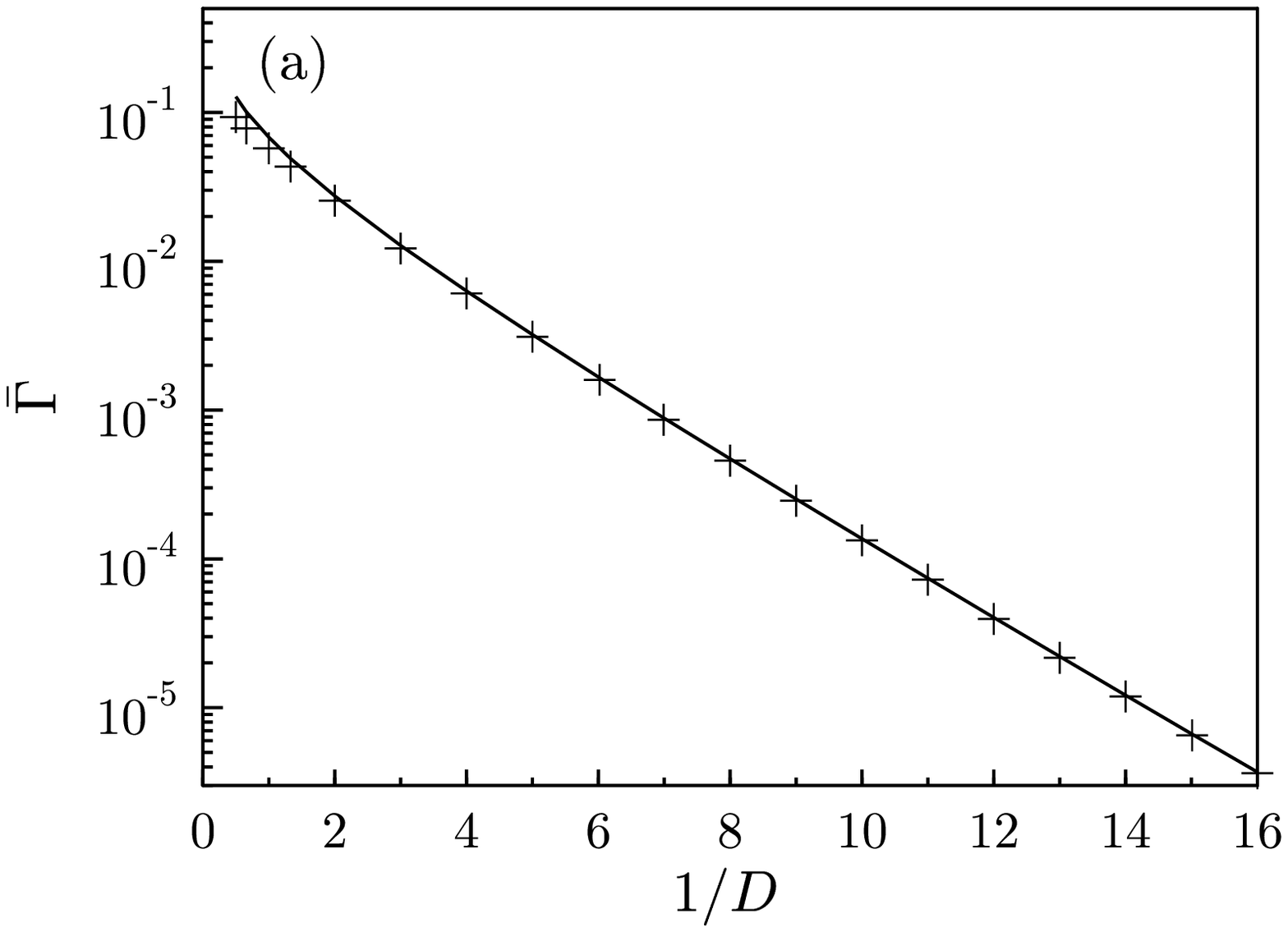,width=1.1\linewidth}

    \vspace*{-7.5cm}
    \epsfig{file=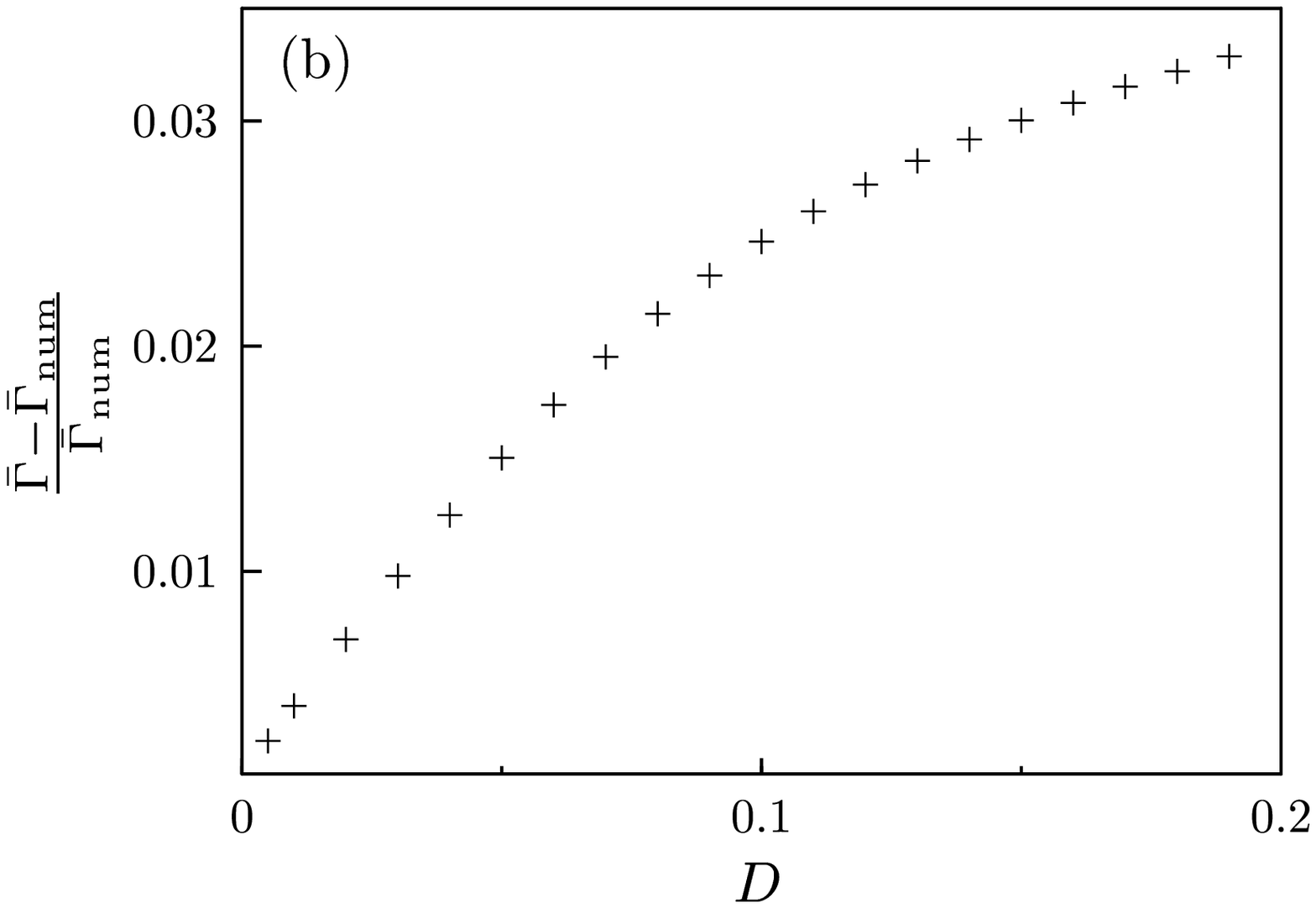,width=1.1\linewidth} \vspace*{-4.2cm}
  \end{center}              
  \caption{(a) Arrhenius plot of the time-averaged rate $\bar\Gamma$ for the
    piecewise linear force field (\ref{5.4}) with the same parameters
    as in fig.4.
    Solid line: Analytical prediction (\ref{4.60},\ref{5.20},\ref{5.25}) by
    neglecting the $\Ord (D^{\gamma})$-term in (\ref{4.60}).
    Crosses: High-precision 
    numerical results, obtained as described in Sect.~V.B.
    (b) Relative difference between
    analytical ($\bar\Gamma$) and numerical ($\bar\Gamma_{\rm num}$) rate.
    }
\end{figure}

\begin{figure}[htbp]
  \begin{center}
    \vspace*{-4.8cm} \epsfig{file=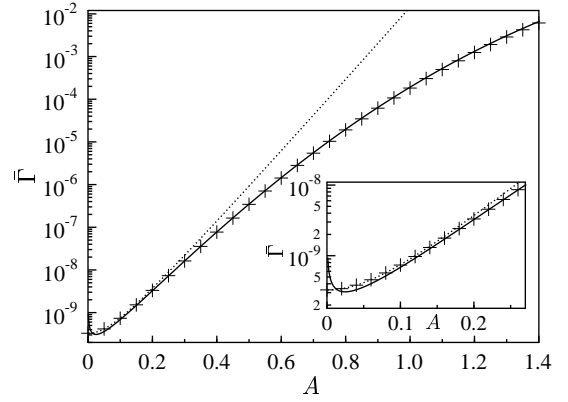,width=1.1\linewidth} \vspace*{-4.3cm}
  \end{center}              
  \caption{Time-averaged rate $\bar\Gamma$ versus driving amplitude $A$ for the
    piecewise linear force field (\ref{5.4}) with parameters
    $x_{s}=\lambda_{s}=-1$, $x_{u}=\lambda_{u}=1$, $\Omega=1$, and $D=0.05$.
    Solid line: Analytical prediction (\ref{4.60},\ref{5.20},\ref{5.25}) by
    neglecting the $\Ord (D^{\gamma})$-term in (\ref{4.60}).
    Dotted line: Theoretical approximation (\ref{5.26})-(\ref{5.29})
    according to Ref.~[23] %\cite{sme99}.
    Crosses: High-precision 
    numerical results, obtained as described in Sect.~V.B.
    Inset: Magnification of the small-$A$ regime.
    }
\end{figure}

\begin{figure}[htbp]
  \begin{center}
    \vspace*{-4.8cm} \epsfig{file=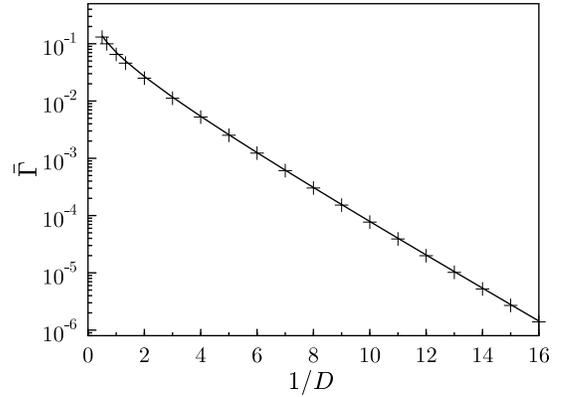,width=1.1\linewidth} \vspace*{-4.3cm}
  \end{center}              
  \caption{Arrhenius plot of the time-averaged rate $\bar\Gamma$ for the 
    cubic potential (\ref{2.5},\ref{5.30}) with parameters $a=1/\sqrt{6}$, $b=1$, $A=0.5$, 
    and $\Omega=1$,
    corresponding to a static ($A=0$) potential barrier $\Delta V = 1$ 
    in (\ref{5.33}) and curvatures $|V''(\xsu)|=1$ in (\ref{5.32}).
    Solid line: Analytical prediction from (\ref{4.60}) without the 
    $\Ord (D^{\gamma})$-term by adopting the calculational procedure from Sect.~IV.H.
    Crosses: High-precision 
    numerical results, obtained as described in Sect.~V.B.
    }
\end{figure}

\end{document}